\documentclass[acmsmall]{acmart}

\usepackage{tikz}
\usepackage{amsmath}
\PassOptionsToPackage{hyphens}{url}
\usepackage{url}
\usepackage{hyperref}
\PassOptionsToPackage{table}{xcolor}
\usepackage{booktabs}
\usepackage{cleveref}
\usepackage{lipsum} 
\usepackage{caption}
\usepackage{subcaption}
\usepackage[inline]{enumitem}
\usepackage{colortbl}
\usepackage{changepage}
\usepackage[acronym]{glossaries}
\usepackage{doi}   
\usepackage{booktabs, multirow} 
\usepackage{adjustbox}

\newcommand{\rev}[1]{{\color{black} #1}}
\newcommand{\mr}[1]{{\color{black} #1}}

\usepackage[inline,nomargin,index]{fixme} 
\fxsetup{status=draft,theme=color,mode=multiuser,inlineface=\itshape,envface=\itshape} 
\FXRegisterAuthor{yz}{ayz}{\color{purple}Yixin}

\definecolor{LightGray}{rgb}{192,192,192}

\hypersetup{colorlinks=true,linkcolor=,urlcolor=black} 

\AtBeginDocument{%
  \providecommand\BibTeX{{%
    \normalfont B\kern-0.5em{\scshape i\kern-0.25em b}\kern-0.8em\TeX}}}

\setcopyright{acmcopyright}
\copyrightyear{2024}
\acmYear{2024}
\acmDOI{10.1145/1122445.1122456}

\acmJournal{TOCHI}
\acmVolume{1}
\acmNumber{1}
\acmArticle{1}
\acmMonth{4}

\begin{document}

\date{}

\title{Nudging Users to Change Breached Passwords Using the Protection Motivation Theory}

\author{Yixin Zou}
\authornote{Work primarily done when the author was a Ph.D. student/postdoc at the University of Michigan.}
\email{yixin.zou@mpi-sp.org}
\affiliation{%
  \institution{Max Planck Institute for Security and Privacy}
  \country{Germany}
}

\author{Khue Le}
\email{khuele@umich.edu}
\affiliation{%
  \institution{University of Michigan}
  \country{USA}
}

\author{Peter Mayer}
\email{mayer@imada.sdu.dk}
\affiliation{%
  \institution{University of Southern Denmark}
  \country{Denmark}
}

\author{Alessandro Acquisti}
\email{acquisti@andrew.cmu.edu}
\affiliation{%
  \institution{Carnegie Mellon University}
  \country{USA}
}

\author{Adam J. Aviv}
\email{aaviv@gwu.edu}
\affiliation{%
  \institution{The George Washington University}
  \country{USA}
}

\author{Florian Schaub}
\email{fschaub@umich.edu}
\affiliation{%
  \institution{University of Michigan}
  \country{USA}
}

\begin{abstract}


We draw on the Protection Motivation Theory (PMT) to design nudges that encourage users to change breached passwords. 
\mr{Our online experiment ($n$=$1,386$) compared the effectiveness of a threat appeal (highlighting negative consequences of breached passwords) and a coping appeal (providing instructions on how to change the breached password) in a 2x2 factorial design.}
\mr{Compared to the control condition, participants receiving the threat appeal were more likely to intend to change their passwords, and participants receiving both appeals were more likely to end up changing their passwords; both comparisons have a small effect size.}
\rev{Participants' password change behaviors are further associated with other factors such as their security attitudes (SA-6) and time passed since the breach, suggesting that PMT-based nudges are useful but insufficient to fully motivate users to change their passwords.}
Our study contributes to PMT's application in security research and provides concrete design implications for improving compromised credential notifications.

\end{abstract}



\maketitle


\section{Introduction}

As digital connections permeate our lives, data breaches --- the unauthorized exposure, disclosure, or loss of personal information --- keep increasing~\cite{solove2022breached}. The types of data compromised range from personal identifiers (e.g., social security numbers in the United States)
to personal health information, bank accounts, emails, and passwords~\cite{itrc2022annual}. The website \texttt{haveibeenpwned.com} has documented billions of compromised account credentials due to data breaches, including those involving high-profile service providers such as Yahoo!, LinkedIn, and Dropbox~\cite{hibp2022websites}. These stolen credentials often end up on the dark web and become trading assets to cybercriminals~\cite{thomas2017data}. With the stolen credentials, attackers can try to break into online accounts at scale through automated login requests (i.e., credential stuffing), meaning that the breach of one service provider's password database can put other accounts using the same or similar passwords at risk~\cite{wang2016targeted}. Once accounts have been compromised, \rev{they can be used for further malicious activity such as distributing malware, impersonation scams, and identity theft~\cite{onaolapo2016happens}.}



Prior work has examined people's reactions to data breaches~\cite{zou2018consumers,mayer2021now,karunakaran2018data} and password-related behaviors~\cite{das2014tangled,wash2016understanding,golla2018site}. The findings suggest that users underestimate the potential harms of data breaches~\cite{zou2018consumers}, reuse a majority of their passwords across sites~\cite{das2014tangled}, and exhibit misconceptions about password security~\cite{golla2018site,ur2016users}. Multiple studies estimate that only between one-quarter to half of all users end up modifying their passwords following data breaches or upon receiving password reset notifications~\cite{golla2018site,huh2017linkedin,thomas2019protecting}.

The low frequency of password changes, compounded by the severity of password attacks, calls for novel approaches to more effectively encourage users to change their passwords after being affected by a data breach. \rev{We see the promise of applying  Protection Motivation Theory (PMT) to achieve this goal.}
In a nutshell, PMT posits that individuals form protection motivation by assessing the threat itself and corresponding coping strategies. \rev{Past research has provided empirical evidence that consumers' limited responses to data breaches could be attributed to low threat perceptions, e.g., thinking that they are less likely to be affected than others~\cite{zou2018consumers,hinds2020wouldn} or considering the breach would have minimal impacts on their lives~\cite{mayer2021now}. Other work has shed light on coping-related impediments that prevent consumers from taking specific actions, such as financial costs~\cite{zou2018consumers,mayer2023awareness} and the action being tiresome and tedious~\cite{hinds2020wouldn}.}
Drawing on PMT, we explore the design space of more effective breach notifications via an online experiment ($n$$=$$1,386$). We assess the effectiveness of PMT-grounded nudges on individuals' password change behavior 
by measuring participants' self-reported intentions and actions. We randomly assigned participants to one of three treatment (nudge) conditions --- a threat appeal, a coping appeal, or both --- in addition to a control condition in which changing the breached password was recommended but no appeal was included. Conducting ecologically valid data breach notification experiments has been a key challenge due to the difficulty of ethically simulating a data breach in research settings and the legal implications when partnering with companies to send out experimental notifications. We manage to achieve an ecologically valid evaluation by situating our nudges in real-world breaches recorded by Have I Been Pwned (HIBP), thereby nudging individuals to take action for passwords that had indeed been compromised in data breaches and subjected them to further risks.
 
We find that compared to the control condition, the threat appeal alone made participants
1.48 times more likely to report the intention to change their passwords; when presented together, the threat and coping appeals made participants 1.54 times more likely to report having changed their passwords. 
We did not find any statistically significant difference between the three treatment conditions. 
Through regression analyses, we identify how password change intentions and actions are also associated with other covariates related to user characteristics (e.g., security attitudes and demographics) and contextual factors (e.g., whether the password is reused and time passed since the breach occurred), suggesting that PMT alone is useful but insufficient to fully \rev{motivate} users to change their passwords. By analyzing participants' open-ended responses, we further identify numerous challenges they experienced when trying to change their passwords, such as forgetting the password that needs to be changed or finding out that they do not have an account with the breached site. Taken together, our findings show the promise of PMT-grounded nudges for motivating password change behavior while contributing new perspectives on inaction: not changing passwords after data breaches could be reasonable when users take alternative actions that still protect them, and is inevitable when there are fundamental issues with the password ecosystem.


\section{Background and Related Work}

We review related work on the Protection Motivation Theory, which serves as the theoretical foundation for this work. We then review prior literature on password-related behaviors (including but not limited to those after data breaches), as they are closely related to the context in which we deployed our nudges.

\subsection{Protection Motivation Theory}

\paragraph{PMT overview.} Protection Motivation Theory (referred to as ``PMT'' onward) seeks to explain individuals' cognitive responses in the face of fear through two appraisal processes, one focusing on the threat itself (\textit{threat appraisal}) and the other focusing on an individual's ability to act against the threat (\textit{coping appraisal})~\cite{rogers1975protection,rogers1983cognitive}. In threat appraisal, people consider the negative consequences of the threat if it occurs (\textit{threat severity}) and how susceptible they are to the threat personally (\textit{threat vulnerability})~\cite{rogers1975protection}. In coping appraisal, people evaluate whether taking a recommended action will mitigate the threat (\textit{response efficacy})~\cite{rogers1975protection}, their confidence in carrying out the action (\textit{self-efficacy})~\cite{weinstein1993testing}, and barriers to taking action such as financial costs, time, and effort (\textit{response costs})~\cite{floyd2000meta}. Higher perceptions of threat severity, threat vulnerability, response efficacy, and self-efficacy increase protection motivation, whereas higher perceived response costs decrease protection motivation~\cite{rogers1983cognitive}.

\mr{PMT was originally developed by Rogers~\cite{rogers1975protection} in the 1970s,  and was revised in the 1980s to include reward and self-efficacy components~\cite{rogers1983cognitive,maddux1983protection}. While PMT has been applied to a wide range of domains, the bulk of its applications is on health-related issues such as cancer prevention, exercise, and smoking~\cite{floyd2000meta}. Within the health domain, there are other theories describing health-related behaviors, such as the health belief model and the theory of reasoned action~\cite{weinstein1993testing}; these theories similarly suggest that motivation toward protection is driven by the desire to avoid potential negative outcomes in the face of  a perceived threat. The transactional model of stress and coping (TSC) also shares similarities with PMT by modeling how individuals evaluate a situation's relevance to their personal well-being (primary appraisal) and options to manage the situation (secondary appraisal)~\cite{lazarus1984stress}.}

\paragraph{PMT in security and privacy research.}
\mr{The majority of PMT's application to the information security and privacy domain has occurred in enterprise settings}, such as using PMT to explain employees' computer security behaviors and adherence to organizational security policies~\cite{burns2017examining,crossler2014understanding,herath2009protection,ifinedo2012understanding,ng2009studying,siponen2014employees,workman2008security}. Other studies have examined PMT \rev{in relation to individual end-users} in interactions with data backups~\cite{boss2015systems}, home computers~\cite{anderson2010practicing,hanus2016impact}, smartphone locking~\cite{albayram2017better,qahtani2018effectiveness}, passwords~\cite{vance2013enhancing}, mobile payment apps~\cite{story2020intent}, Tor browser~\cite{story2022increasing}, and more. 
\mr{Researchers often incorporate other behavioral change theories in addition to PMT to explain security behaviors, highlighting the role of factors such as} personal responsibility~\cite{boehmer2015determinants,larose2008promoting,shillair2015online}, attitudes~\cite{siponen2014employees}, social norms~\cite{siponen2014employees,herath2009protection,johnston2010fear}, psychological capital~\cite{burns2017examining}, and prior negative experience~\cite{lee2008keeping} \mr{in addition to threat and coping appraisals}.
  
PMT can also be applied to develop nudges that promote security behaviors --- which is the goal of our study. The majority of existing work on PMT-based nudges has compared PMT condition (combining threat and coping appeals) versus a control condition~\cite{boss2015systems,jenkins2014improving,albayram2017better,story2020intent,story2022increasing},
proving that PMT interventions promote more secure behaviors consistently across contexts. 
The result may indicate that increasing protection motivation requires both threat and coping appeals, but to prove this argument, we need to examine the effects of threat and coping appeals in isolation, and prior work has provided relatively limited insights on this. Our research takes inspiration from van Bavel et al.'s study, in which the authors separated threat and coping appeals in examining their effects on four behaviors related to online security, finding
that the coping appeal was as effective as both appeals combined, but not so the threat appeal alone~\cite{van2019using}. 

Nevertheless, prior work has provided inconclusive results on which one is more effective between threat and coping appeals. Some studies have identified coping appraisal as a significant predictor of security intention or behavior~\cite{boehmer2015determinants,burns2017examining,crossler2014understanding,hanus2016impact,herath2009protection,ifinedo2012understanding,johnston2010fear,larose2008promoting,lee2008keeping,ng2009studying,siponen2014employees}, but among these studies many also found significant effects for threat-related constructs, either threat severity or vulnerability alone~\cite{burns2017examining,herath2009protection,ifinedo2012understanding,johnston2010fear,lee2008keeping} or both~\cite{siponen2014employees,anderson2010practicing}. Other studies suggest that threat perceptions did not directly impact intention or behavior, but moderated the effect of coping-related constructs~\cite{ng2009studying} or had interaction effects with the subject's occupational background~\cite{crossler2014understanding}. Mayer et al. suggested that threat vulnerability and rewards of maladaptive response are not reliable predictors of security intention in their literature review, but also cautioned that the finding was based on a limited set of studies~\cite{mayer2017reliable}. \rev{A critical underlying assumption of the threat dimension is that the threat holds personal relevance to the message recipient, which is not always the case in enterprise security; threats such as password theft and data loss are relevant to an organization's information assets, but not necessarily a particular employee. As such, Johnston et al. found that an enhanced fear appeal by adding the dimension of personal relevance can more effectively influence compliance intentions~\cite{johnston2015enhanced}.}




\paragraph {A possible intention-behavior gap.} Behavioral intention is often used as a proxy for protection motivation and is considered a reasonable predictor of behavior according to the Theory of Planned Behavior~\cite{ajzen1991theory}. However, prior work has revealed a possible intention-behavior gap, calling for measuring both variables when possible. For example, a meta-analysis of experimental evidence showed that a medium-to-large-sized change in intentions led to only a small-to-medium-sized change in behavior~\cite{webb2006does}. Key challenges that people may encounter as they strive to enact their intentions include failing to get started, failing to keep goal pursuit on track, and failing to bring goal pursuit to a successful close~\cite{sheeran2016intention}. Correspondingly, there are self-regulatory mechanisms that target these challenges to help people realize their intentions, such as forming implementation intentions~\cite{gollwitzer1999implementation} and monitoring goal progress~\cite{harkin2016does}.

Among the limited number of studies that examined the intention-behavior gap in security contexts, Crossler et al. found that the costs of implementation (e.g., time and inconvenience) could be a strong deterrent to full compliance for employees to follow Bring Your Own Device policies~\cite{crossler2014understanding}. Similarly, Jenkins et al. found that users' desire to minimize required effort negatively moderates the relationship between positive intentions and actual security behavior~\cite{jenkins2021mitigating}. \rev{On PMT-based nudges for using secure mobile payments, Story et al. measured intention and behavior with a one-week gap in between, finding that their PMT nudges as well as implementation intention-based nudges can help participants translate intention into actual behavior~\cite{story2020intent}. However, none of these studies are designed in ways that allow investigating whether PMT-based nudges have triggered permanent and habitual rather than one-time behavioral changes among users. Our study's set up is similar to that of Story et al.~\cite{story2020intent} to measure the intention-behavior gap, but a possible direction for future work is to adopt a more longitudinal evaluation of the actual behavior.}

%

Altogether, our study contributes to the PMT literature by (1) applying PMT to encourage protection motivation in a new context, i.e., changing passwords after data breaches; (2) separating threat and coping appeals to understand their relative effectiveness; and (3) measuring both intention and behavior to get a complete picture of protection motivation. 


\subsection{Password Policies, Behaviors, and Nudges}

\paragraph{Password policies.} Service providers use password composition policies to prevent users from creating easily guessed passwords~\cite{shay2016designing}, as strong passwords, measured by length and complexity, are harder to crack than shorter and simpler passwords~\cite{nist-17-sp800-63-3}. \rev{However, there are inconsistencies in the provided guidelines, e.g., the eight-character minimum length requirement by the US National Institute of Standards and Technology~\cite{nist-17-sp800-63-3} versus the recommendation of using three random words by the UK National Cyber Security Center~\cite{ncsc2018password}.}
Many service providers go above and beyond to require different character classes to be included or periodic password changes~\cite{cranor2016time}, but research has shown that requiring longer passwords with fewer composition requirements provides more security and usability benefits~\cite{shay2014my} and users have been found to circumvent the character class requirements in predictable ways~\cite{ur2015added}.


Research of password corpora has also identified characteristics of common passwords, which inform the creation of password policies about certain phrases to avoid. According to an analysis of leaked passwords from RockYou, the most popular passwords were ``123456,'' ``password'' and ``iloveyou''~\cite{vance2010if}. Other prevalent semantic themes in passwords include names, locations, dates, animals, and money~\cite{veras2014semantic,medlin2007empirical}.

We reconcile inconsistencies in existing password requirements and identify the common grounds: a good password policy should require a minimal length (eight characters according to the NIST~\cite{nist-17-sp800-63-3} or longer if the organization prefers) and some but not too many classes of special characters, highlight common phrases to avoid, and remove arbitrary password expiration periods. We incorporated such information as well as common passwords to avoid when developing our coping appeal.

\paragraph{User behaviors.}

Despite good-faith efforts in protecting the security of their personal information, most users struggle to comply with password policies~\cite{shay2014my} and create weak passwords~\cite{ur2015added,pearman2017let,habib2018user}. Password reuse is prevalent~\cite{das2014tangled,florencio2007large,pearman2017let}: Das et al. estimated that 43-51\% of users reuse the same password across multiple sites in their study~\cite{das2014tangled}, and Pearman et al. found that the rate would be higher when taking partial reuse into account~\cite{pearman2017let}. Password reuse becomes more likely as more passwords are created or when the reused password is short and simple to memorize~\cite{gaw2006password}. Users match password strength to the account's relative importance~\cite{ur2015added} and rarely change their passwords unless they forget about the password~\cite{stobert2014password}.

Even with external stimuli, users do not always comply with password change advice. Thomas et al.'s work on breached credentials found that only 26\% of the warnings resulted in users migrating to a new password~\cite{thomas2019protecting}. In Bhagavatula et al.'s study based on real-world password data, only 33\% of all participants who had accounts on the breached site changed their passwords, and only 13\% did so within three months of the breach announcement~\cite{bhagavatula2020how}. In a case study of LinkedIn, Huh et al. found that less than half of participants changed their LinkedIn password upon receiving the password reset notification from the company~\cite{huh2017linkedin}. For users who change the breached passwords, they rarely change the same or similar passwords on other sites~\cite{bhagavatula2020how}, and their new passwords tend to be similar to their old ones~\cite{golla2018site}. Trust issues can also arise with third-party advice providers. In studying users' perceptions of Chrome's compromised credential notifications, Huang et al. found that some participants falsely assumed Chrome learns about users' plain-text credentials or expressed privacy concerns about Google's management of users' data~\cite{huang2022users}.

Methodologically, prior research has measured password behaviors by tracking participants' log data~\cite{florencio2007large,bhagavatula2020how,forget2016user}, analyzing passwords from public or private datasets~\cite{das2014tangled,mazurek2013measuring}, observing password creation in-situ~\cite{ur2015added}, and relying on participants' self-reported data~\cite{wash2017can,golla2018site,huh2017linkedin}. We evaluate our nudges based on participants' self-reported data, which could still be prone to social desirability and recall biases. However, our method is grounded in high ecological validity as we present participants with real-world breaches that affect their email addresses and potentially other personal information. 


\paragraph{Helping users with passwords.}

Password managers are tools that combine secure password storage and retrieval with random password generation to help users deploy strong, unique passwords without memorability issues~\cite{pearman2019why}. However, users are often uncertain about what
password managers are, how to use them, and whether they
are trustworthy~\cite{alkaldi2016people,stobert2014password}. Users may believe there is little to protect, worry about having a single point of failure, or have prior negative experiences with password managers~\cite{pearman2019why,ray2021older}. Alkaldi et al. studied how to encourage the adoption of password managers by satisfying users' self-determination needs in terms of autonomy, relatedness, and competence~\cite{alkaldi2019encouraging}. 

Other research has sought to develop nudges that help users create stronger passwords \rev{in general. The nudge can be} visual or text indicators that provide feedback on the password's strength~\cite{carnavalet2015large,ur2012does}, an estimate of how long it would take to crack the password~\cite{wheeler2016zxcvbn,vance2013enhancing}, 
or social nudges that compare the strength of the user's password with others~\cite{egelman2013does}. Peer et al. showed that password nudges could be more effective when they are also personalized to individuals' decision-making style~\cite{peer2020nudge}. 

Our study expands prior literature by examining the applicability of PMT in encouraging better password behaviors and grounding our nudges in an environment with high ecological validity, in which participants act on their own passwords and accounts based on notifications of real-world breaches. 
\section{Method}

Between July and August 2022, we conducted an online experiment to evaluate the effectiveness of PMT-based nudges in encouraging people to change passwords after data breaches. Our study was approved by the Institutional Review Boards at the University of Michigan and the George Washington University. 

To increase the ecological validity of our evaluation, we drew on Mayer and Zou et al.'s methodology~\cite{mayer2021now} and pulled participants' breach records from Have I Been Pwned (HIBP) using its public API, thereby situating our nudges in real-world breaches known to affect individual participants. HIBP is a database maintained by security expert Troy Hunt who routinely analyzes password dumps and text storage sites on the Internet to collect information about leaked account credentials. As of August 2022 (when we finished data collection), the site listed 622 breached websites with over 11 billion breached accounts.

\subsection{Study Design}

The purpose of our study is to evaluate to what extent PMT-based nudges are effective at encouraging consumers to change their passwords after being affected by a data breach. The key dependent variables are participants' self-reported password-change intentions and behaviors. 
Because prior work has not provided conclusive evidence regarding the relative importance of threat and coping appraisals, our experiment examines the effect of threat and coping appeals together and in isolation.

In total, our experiment has four conditions:

\begin{itemize}

    \item Control: Participants were not presented with any threat or coping-related information. As a baseline, participants were still presented with a prompt: ``We recommend that you change the password for your [site name] account'' as well as information about a data breach that affected them. This baseline prompt appeared in all conditions.

    \item Threat-only: Participants were presented with information about the risks associated with passwords being leaked in a data breach (threat severity) and how the risks could affect them personally (threat vulnerability), cf. Figure~\ref{fig:threat-text}.
    
    \item Coping-only: Participants were presented with information about how to change their password (self-efficacy), how changing the password reduces the threat (response efficacy), and estimated time to change their password (response costs), cf. Figure~\ref{fig:coping-text}.
    
    \item Threat$+$coping: Participants were presented with the combined information about both threat and coping described above.

\end{itemize}

\begin{figure}[t]
\fbox{\parbox{.97\linewidth}{
  \small
 \textbf{What are the risks}
    \begin{itemize}[noitemsep,topsep=0pt]
        \item Criminals may access your account to steal your personal information, impersonate you, or make fraudulent purchases in your name.
        \item If you used the same password elsewhere, criminals may take over your other accounts too.
        \item Criminals use automated programs to test compromised passwords on hundreds of accounts in just a few seconds. You’re at risk regardless of whether you are a promising target or not.
        \item Once your password is out there, criminals may try to take over your account anytime after a breach, no matter how long ago the breach happened.
    \end{itemize}
}}
\caption{The threat appeal we used in our study.}
\label{fig:threat-text}
\vspace{1em}
\fbox{\parbox{.97\linewidth}{
    \small
    \textbf{How to change your password}
    Changing your [site name] account password would prevent criminals from using the breached password to access your account. It only takes a few minutes. Just follow these easy steps:
    \begin{enumerate}[noitemsep,topsep=0pt]
        \item \textbf{Go to [site URL] and log into your account.}
        \begin{sloppypar}
        Unsure if you have a [site name] account or can’t log into it? Contact [site name] to recover the account or have your account deleted. You can usually find contact information in the privacy policy.
        \end{sloppypar}
        \item \textbf{Create a unique and strong password in account settings.}
        \begin{sloppypar}
        Longer passwords are best. Do not reuse the same password for other accounts. Check out 
        \href{https://www.cisa.gov/uscert/ncas/tips/ST04-002}{this guideline}~\cite{cisa2019guideline}
        for more do's and don'ts about passwords.
        \end{sloppypar}
        \item \textbf{You're all set!}
        \begin{sloppypar}
        If you used your old password for other accounts, make sure to change your password for those accounts too.
        \end{sloppypar}
    \end{enumerate}
}}
\caption{The coping appeal we used in our study.}
\label{fig:coping-text}
\end{figure}

We drew on existing literature and guidelines for the text and design of our nudges. The threat appeal (1) features negative things that could happen to affected users when a data breach involves login credentials~\cite{golla2018site,thomas2017data} (matching \textit{threat severity}) and (2) addresses optimism bias and hyperbolic discounting that people may experience after data breaches~\cite{zou2018consumers,karunakaran2018data} (matching \textit{threat vulnerability}). The coping appeal (1) builds the connection between changing the password and reducing the chance of account compromise (matching \textit{response efficacy}), (2) gives a list of concrete steps to take, including a URL to the breached site and guidelines about how to create strong passwords~\cite{cisa2019guideline} (matching \textit{self-efficacy}), and (3) highlights that the effort ``only takes a few minutes'' and is ``easy'' (matching \textit{response costs}). In line with prior work's recommendations~\cite{huang2022users,emami2019exploring,cranor2012necessary}, we followed a layered approach --- presenting the text in multiple interactive expandable boxes with a short delay between boxes --- to reduce participants' cognitive load while nudging them to pay attention. Participants would have to expand all text and see the full intervention before being able to proceed to the next page. 

\begin{figure}
  \centering
  \includegraphics[width=0.8\linewidth]{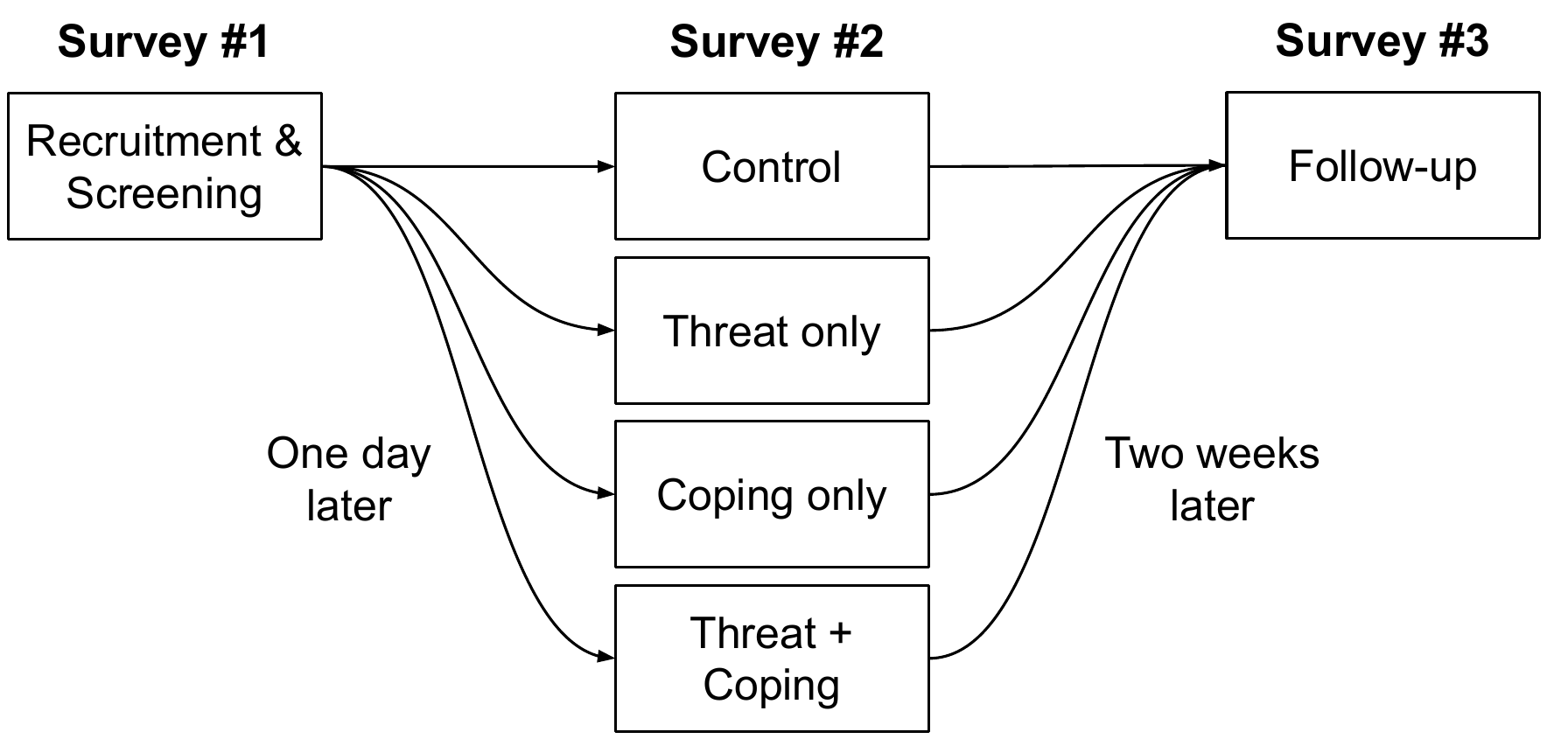}
    \caption{An overview of our experiment's procedure.}
    \label{fig:study-design}
\end{figure}

The experiment was carried out through three online surveys (see Figure~\ref{fig:study-design} for an overview and Appendix~\ref{app:pmt-survey-material} for full survey questionnaires). 
In the screening survey, we obtained participants' consent and asked them to provide an email address to be queried in HIBP for all breach records associated with that email address. We determined each participant's eligibility based on the query results and invited eligible participants back for the main survey. In the main survey, participants were presented with different nudging text based on the condition they had been randomly assigned to and were asked to report their password change intention. For participants who indicated the password had not been changed since the breach (Q\ref{q:pmt-intention-close-ended}) --- meaning that the password needs to be changed --- we invited them back two weeks later for a follow-up survey that measured whether they had changed the password after receiving the nudge. 


\subsection{Protocol}

\paragraph{Pilot testing.} To ensure that the nudges we developed worked as intended
and that the questionnaire was understandable, the first author conducted four cognitive walkthrough sessions~\cite{lazar2017research} with participants recruited via their social networks. Each session lasted about 60 minutes, during which participants were asked to take all three surveys while verbalizing their thoughts and questions throughout (see Appendix~\ref{app:walkthrough-protocol} for the cognitive walkthrough protocol). All four participants confirmed that the text was intuitive and aligned with their understanding of threats related to breached passwords as well as challenges regarding changing passwords. Based on the feedback we received, we made minor tweaks to the questionnaire, such as including social login~\cite{flores2020social} as an option for creating accounts (i.e., choosing ``sign in with Apple'' or ``sign in with Google'' rather than creating one's own username and password).

We conducted a second, larger round of pilot testing with participants recruited via Prolific, a crowdsourcing platform commonly used in social science and behavioral research. We collected 107, 76, and 69 complete responses for the screening, main, and follow-up surveys, respectively. Each participant was compensated \$1.00, \$3.00, and \$1.60 for completing each survey respectively. Based on the pilot data, we made further changes to the questionnaire design and compensation.
We also used the pilot data to determine participants' compensation and effect sizes to be considered in power analysis (more in Section~\ref{subsec:pmt-analysis}).

\paragraph{Recruitment \& Data Collection.}

We recruited participants for the screening survey via Prolific. We made our screening survey open to prospective participants who spoke English, were at least 18 years old at the time of the survey, and were living in the United States. We also limited the survey to participants who could take it in Chrome/Firefox on a desktop computer, since we found through pilot testing that our survey infrastructure ran into occasional technical issues in other environments.

We began collecting data for the screening survey in July 2022 and concluded data collection for the follow-up survey in August 2022. We obtained 2,412 complete responses to the screening survey. In creating our sample, we sought a diverse representation of gender, age, and educational attainment. Specifically, we used the ``balanced sample'' feature on Prolific, which allows us to distribute the study evenly to male and female participants.\footnote{While Prolific uses sex as one of its pre-screener variables, we used gender in our questionnaire, also including ``non-binary,'' ``prefer to self-describe,'' and ``prefer not to say'' options.} Furthermore, we released slots for the screening survey in small batches ($n$$\approx$$100$), monitored the demographic distributions of incoming data, and used Prolific's pre-screeners to balance age and education as needed.\footnote{Since we found that random sampling on Prolific tends to attract younger and more educated participants, we targeted the screening survey to middle-aged ($>=$35 years old) and older ($>=$45 years old) participants in nine out of 24 batches and people who do not have a college degree in two out of 24 batches. We only did such stratification for the screening survey since the main and follow-up survey participants came from the screening survey.}

The actual experiment began with the main survey and continued through the follow-up survey. Among participants who completed the screening survey, 1,824 (75.6\%) were deemed eligible. We invited these participants back for the main survey one day after they completed the screening survey. Among the 1,654 (90.7\%) participants who returned and completed the main survey, 1,388 (83.9\%) selected ``yes'' or ``no'' for the password change intention question, and the remaining 266 (16.1\%) selected ``already changed.''\footnote{Participants who selected ``already changed the password'' for the intention question were almost evenly distributed across the four conditions (X: 61/22.9\%; T: 73/27.4\%; C: 69/26.0\%; CT: 63/23.7\%).} We excluded participants who selected ``already changed'' from the follow-up survey and from our data analysis since the nudges would not apply to them as there is no need for them to change their password again. Among participants who received the follow-up survey invitations, 1,176 (84.7\%) returned and completed the follow-up survey.

Our pilot data suggested that participants on average took three, nine, and four minutes to complete the screening, main, and follow-up surveys respectively. Aiming to compensate participants at least \$15/hour, we adjust the compensation accordingly: \$0.80 for the screening survey, \$2.40 for the main survey, and \$1.00 for the follow-up survey. The actual median completion times were 1.84 minutes (screening), 6.64 minutes (main), and 3.71 minutes (follow-up), corresponding to a rate of \$26.08/hour, \$21.69/hour, and \$18.76/hour respectively.



\paragraph{Screening Survey.} 

After obtaining their informed consent, we asked participants to provide an email address, which we used to query the HIBP database and obtain information about data breaches in which the participant's email address had been exposed. We informed participants that the querying task would be completed in a privacy-preserving manner: by keeping participants' email addresses in ephemeral storage and deleting them after the query had been sent, we never had direct exposure to participants' email addresses, nor did we store them anywhere or include them in our analysis. Participants who were uncomfortable with providing their email addresses could opt out. \rev{Among the 2,625 participants who consented to participate in the study, 172 (6.6\%) opted out when asked to provide an email address.}


For participants who provided an email address, we completed the automated HIBP query and informed them of their eligibility for our experiment based on the query results. A participant would be 
eligible if they provided an email address for their own or shared email account, rather than an email address that belongs to someone else entirely or is made up for the study. Furthermore, the participant needed to have at least one valid breach for their provided email address. We considered that a breach is valid (1) if passwords were among the types of breached information, which makes changing the password a reasonable option, and (2) changing the password is indeed a viable option on the breached site, meaning that the site should be functioning, have an account creation feature, and conduct business with average consumers rather than other businesses.\footnote{To ensure the second criterion is met, we manually inspected all breached sites listed in the HIBP database in June 2022, right before launching the experiment. Our manual filtering led us to exclude 230 out of the 598 breaches (38.5\%) in the HIBP's database at the time of our study.} Participants who were ineligible for our study were redirected to the final page showing the breach records associated with their provided email addresses. 

For eligible participants, we asked them to indicate how they used the email account and collect demographic information at the end of the screening survey, as these are all possible covariates of one's password change behaviors. For the email-related questions, we drew on the same set used in Mayer and Zou et al.~\cite{mayer2021now}, \rev{including how often they checked this email account (e.g., every day or a few times a week), how long it has been used (free-text responses with fixed time units), and what they used the account for (e.g., professional/personal correspondence or account creation, and participants could select multiple options for this question). For account creation, we further asked participants if they used it to create accounts that are low-value (i.e., accounts that one does not really care), medium-value (e.g., social media), and high-value (e.g., banking).}


\paragraph{Main Survey.}

\begin{figure}
  \centering
  \includegraphics[width=0.75\linewidth]{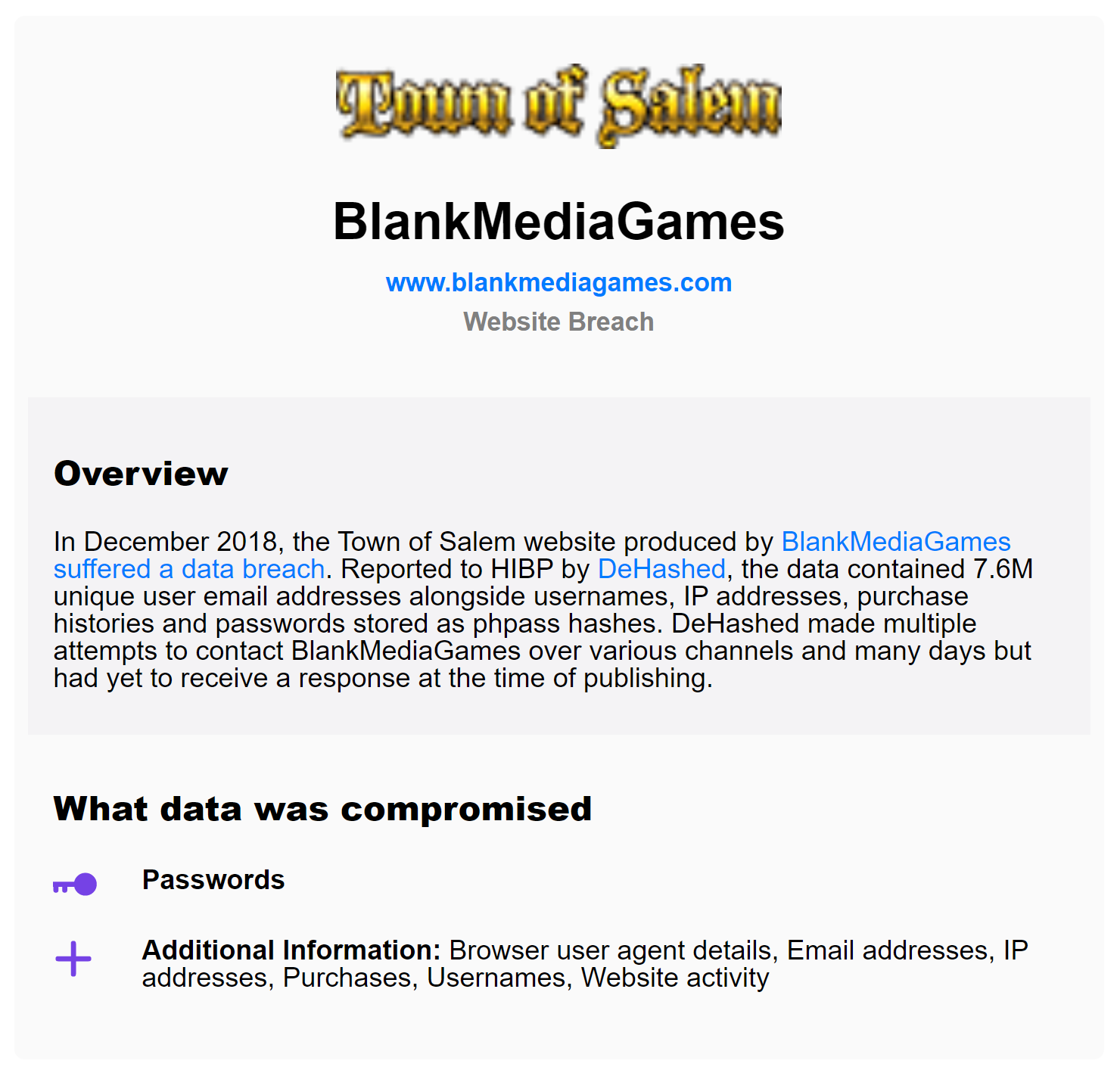}
    \caption{Example breach information shown to participants.}
    \label{fig:breach-featured}
\end{figure}

For participants who passed the screening, we sent invitations to the main survey one day later. The main survey started by showing a breach that involved passwords randomly chosen from their breach records. We provided a short description of the breach, the site's logo and name, and the types of compromised data, all drawing from HIBP. We highlighted passwords being compromised and listed other data types as ``additional information'' (cf. Figure~\ref{fig:breach-featured}). We then asked participants about their awareness in three dimensions: whether they had heard of the site, whether they had known they were affected by the breach, and whether they had an account with the site. For those with an account, we further asked them to specify their account usage, including the account's age, frequency of use, and perceived importance.

Next, we randomly assigned participants to one of the four conditions: all conditions included the password change prompt, but participants in the treatment conditions received additional information targeting threat appraisal, coping appraisal, or both. After showing the nudge, we asked participants to indicate their password change intention (``yes'' ``no'' ``already changed'') and to explain their choice in a text box. We also asked participants whether they had used the breached password for other online accounts.

Following the password reuse question, we asked participants a series of Likert questions to assess five PMT constructs --  threat severity, threat vulnerability, response efficacy, self-efficacy, and response costs -- using scales adapted from prior work for each construct~\cite{story2020intent,boehmer2015determinants}. We used these responses as manipulation checks to gauge how effective our nudges were at targeting our participants' threat and coping appraisals. We also included two attention check questions.
The main survey ended with questions about the participant's security attitudes (measured by SA-6~\cite{faklaris2019self}) and prior negative experiences (adapting questions from Zou et al.~\cite{zou2020examining}) as additional covariates of password change intention and behavior.

\paragraph{Follow-up Survey.}

For participants who were eligible for the follow-up survey, we sent out the invitation two weeks after they completed the main survey. We decided two weeks after as the follow-up time based on pilot testing, as all pilot participants who ended up changing their password did it within two weeks. We reasoned that this timeframe gave participants enough time to change their password if they wanted to do it on certain less busy days (e.g., weekends) but was not too long to cause recall issues. 


The follow-up survey began by reminding participants of the previous surveys they took and the breach featured. Next, we asked participants to describe what they had done in response to learning about the breach in the main survey via an open-ended question, followed by an attention check asking participants to select the name of the featured breach from a list including the correct option and three decoys. We then asked participants to specify whether they had changed their password since taking the main survey, why, and what they did to passwords for other accounts. Those who indicated they had changed their password since taking the main survey were asked to provide more details, such as when they changed the password and what mechanisms they used to remember the new password. 

Participants could further optionally upload a screenshot of the password reset confirmation email. We adopt this approach from~\cite{huh2017linkedin} as a measure to validate participants' self-reported behavior and set a \$1.00 bonus payment for those who uploaded a valid screenshot. \rev{Importantly, the screenshot upload prompt (Q\ref{q:pmt-screenshot-question}) came after the question that asked participants whether they actually changed their password (Q\ref{q:pmt-action-binary}), and participants were not allowed to go back and change their earlier responses. By implementing this sequence, we were able to ensure that participants' reported password change behaviors occurred as a result of our nudge rather than because of the additional monetary incentive from the screenshot upload prompt.}



\subsection{Data Analysis}
\label{subsec:pmt-analysis}

We pre-registered our study protocol, hypotheses, and data analysis plan prior to data collection on Open Science Framework.\footnote{\url{https://doi.org/10.17605/OSF.IO/P49A6}}


\paragraph{Hypotheses.} Our key independent variable is the condition: control (X), threat only (T), coping only (C), and threat and coping combined (CT). Our key dependent variables are participants' intention and self-reported action to change their breached passwords. Because the four conditions had different amounts of threat- and coping-related information, and because the control condition does not have any threat or coping information, we made the following hypotheses for password change intention:

\textbf{H1}: There is a significant association between the condition and whether or not the participant would intend to change their breached passwords.
\begin{itemize}[noitemsep,nolistsep]
    \item \textbf{H1a}: Participants in the control condition will exhibit lower password change intentions than those in the threat-only condition. (\textit{intention: $X$$<$$T$})
    \item \textbf{H1b}: Participants in the control condition will exhibit lower password change intentions than those in the coping-only condition. (\textit{intention: $X$$<$$C$})
    \item \textbf{H1c}: Participants in the control condition will exhibit lower password change intentions than those in the threat$+$coping condition. (\textit{intention: $X$$<$$CT$})
\end{itemize}

Assuming that the nudge's effect on intention in the main survey carries over into the follow-up survey, we made the following hypotheses for password change behavior:

\textbf{H2}: There is a significant association between the condition and whether or not the participant would end up changing their breached passwords.
\begin{itemize}[noitemsep,nolistsep]
    \item \textbf{H2a}: Fewer participants in the control condition will end up changing their breached passwords than those in the threat-only condition. (\textit{behavior: $X$$<$$T$})
    \item \textbf{H2b}: Fewer participants in the control condition will end up changing their breached passwords than those in the coping-only condition. (\textit{behavior: $X$$<$$C$})
    \item \textbf{H2c}: Fewer participants in the control condition will end up changing their breached passwords than those in the threat$+$coping condition. (\textit{behavior: $X$$<$$CT$})
\end{itemize}

We did not plan to apply p-value corrections for the pairwise comparisons since our hypotheses are confirmatory rather than exploratory. In addition, we did not include directional hypotheses regarding differences among the three treatment conditions because prior work has not provided sufficient evidence on whether the threat or coping nudges may be more effective or whether the two nudges combined would be more effective than each presented in isolation~\cite{van2019using,mayer2017reliable}. However, we were still interested in understanding their differences, and we ran additional pairwise comparisons as exploratory analyses to obtain insights. 

\paragraph{Data Cleaning.}

We retained all complete responses for the screening survey. We used the attention checks in the main and follow-up surveys to flag responses that required further examination. Among the 1,388 participants who completed the main survey --- excluding those who already changed passwords --- 32 failed one of the attention checks, and two failed both attention checks. Among the 1,176 participants who completed the follow-up survey, one failed the attention check. We excluded responses from the two participants who failed both attention checks in the main survey and the one who failed the attention check in the follow-up survey. We retained the remaining responses since the data for other parts of the survey was detailed and insightful. We also retained responses from the 211 participants who completed the main survey but did not complete the follow-up survey since they still contributed insights for half of our key hypotheses.  

After data cleaning, the number of complete and valid responses for the main survey was 1,386 --- we regard this as the final sample size of our study. The distribution across conditions is almost even for the main survey (control: 349/25.2\%; threat-only: 339/24.5\%; coping-only: 348/25.1\%; threat$+$coping: 350/25.2\%) and the follow-up survey (control: 291/24.8\%; threat-only: 304/25.9\%; coping-only: 297/25.2\%; threat$+$coping: 283/24.1\%). 

\paragraph{Statistical Analyses.}

To confirm H1 and H2, we conducted two omnibus $\chi^2$ tests: both tests used the condition assigned (four levels) as the independent variable, intention (yes/no, excluding ``already'') and action (yes/no) as the dependent variable, respectively. To confirm H1a-H1c and H2a-H2c, we conducted pairwise $\chi^2$ comparisons to detect the differences between the control and any treatment conditions. Following practices in prior work~\cite{story2020intent}, we planned to run these pairwise comparisons regardless of the omnibus test results since we had solid and theory-informed hypotheses that any treatment would better encourage password changes than the control condition.

We also planned and conducted a series of exploratory analyses. To gather insights into the relative performance of threat versus coping appeals, we conducted pairwise comparisons between the three treatment conditions. To understand to what extent the threat and coping appeals address participants' threat and coping appraisals respectively, we conducted Kruskal-Wallis tests to investigate the relationship between the condition (IV) and each of the five PMT constructs (DV) as manipulation checks. 

We were also interested in knowing whether the effect of the nudge would remain robust after controlling for other covariates, such as individuals' account usage and demographics. To this end, we built logistic regression models with the condition and other covariates as IVs and intention/action as the DV. 
To avoid model fit problems caused by too few observations in a category, we binned the demographic data in Table~\ref{tab:pmt-participants} into fewer categories for the regression analyses: gender (binary: men or women), age (three levels: 18-34, 35-54, 55+), educational attainment (three levels: high school or less, some college, Bachelor's degree or above), and annual household income (three levels: $<$\$50k, \$50-100k, $>$\$100k). 
We report odds ratios, confidence intervals, and p-values for regression results.


\paragraph{Qualitative Analysis.} We analyzed participants' open-ended text responses using thematic coding~\cite{saldana2015coding} to gather deeper insights into their reasoning behind password change intention (Q\ref{q:pmt-intention}), what they did in general after learning about the breach (Q\ref{q:pmt-response}), and reasoning behind password change action (Q\ref{q:pmt-action}). The first author created an initial codebook and sought feedback from the research team in multiple discussions to iteratively improve the codebook. Since our analysis seeks to report quantitative analysis of qualitative data (e.g., that a theme showed up in X\% of our data), the first author then worked with the second author to double-code 20\% of responses to ensure sufficiently high inter-rater reliability~\cite{mcdonald2019reliability}. Cohen's $\kappa$ was consistently above 0.80 across three rounds of comparison, and the final values were 0.83 for Q\ref{q:pmt-intention}, 0.88 for Q\ref{q:pmt-response}, and 0.86 for Q\ref{q:pmt-action}. After that, the first and second authors resolved coding discrepancies through discussions, and the first author went back to review all responses using the finalized codebook (Appendix~\ref{app:codebook}). 



\paragraph{Power Analysis.}

We conducted an \textit{a priori} power analysis to determine our target number of participants based on effect sizes observed in the pilot data. Our pilot data suggested a small-to-medium effect ($w$$=$$0.15$) for the omnibus $\chi^2$ test on intention and a small effect ($w$$=$$0.10$) for pairwise comparisons on intention. We based our power analysis on pairwise comparisons since we wanted to ensure we would have enough data to detect between-condition differences. For 80\% power at $\alpha$$=$$0.05$ and $df$$=$$1$, G*Power suggested we need 1,570 participants in total for the main survey. We did not achieve this goal due to budgetary constraints and an underestimation of participants whose data got excluded from the analysis because they had changed the password already (16\% of all completed main survey responses, whereas our original estimate was 5\%). However, our final sample size still enabled us to detect a small effect for the omnibus $\chi^2$ test ($w$$=$$0.09$ for intention; $w$$=$$0.10$ for action) and a small-to-medium effect for the pairwise comparisons ($w$$=$$0.11$ for intention; $w$$=$$0.12$ for action).
\section{Quantitative Results}


In this section, we summarize participant demographics and breaches featured in our study, present key findings in relation to our hypotheses, then present manipulation check and regression results to contextualize and interpret the differences between conditions. 



\subsection{Sample}
\label{subsec:sample}

\paragraph{Participant profile.} 

\begin{table*}[tp]
\footnotesize
\centering
\begin{tabular}{l l l}
\textbf{Metric} & \textbf{Sample} & \textbf{Census}\\
\cmidrule(lr){1-1} \cmidrule(lr){2-2} \cmidrule(lr){3-3}
Women, Men, Non-binary & 50.9\%, 46.5\%, 1.9\% 
& 51\%, 49\%, n/a\\[2mm]

18-24, 25-34 years& 16.0\%, 22.9\% & 7\%, 14\%\\
35-44, 45-54 years & 21.5\%, 19.5\% & 13\%, 13\%\\
55-64, 65 years or older & 13.3\%, 5.4\% & 13\%, 15\%\\ [2mm]

High school or less, Some college & 11.6\%, 23.7\% & 37\%, 15\%\\
Associate's degree (aca./voc.) & 11.8\% & 11\%\\
Bachelor's degree & 36.9\% & 24\% \\
Advanced degree (Master's/professional/doctoral) & 15.7\% & 14\%\\[2mm]

$<$\$25k, \$25k-\$50k & 16.0\%, 22.3\% & 19\%, 20\% \\
\$50k-\$75k, \$75k-\$100k & 20.1\%, 14.9\% & 16\%, 12\%\\
\$100k-\$150k, $>$\$150k  & 13.5\%, 10.0\% & 14\%, 19\%\\ [2mm]

Asian, Black & 7.8\%, 6.8\% & 6\%, 14\%\\
White, Two or more races & 76.9\%, 5.0\% & 76\%, 3\%\\
Hispanic/Latino  & 9.8\% & 19\%\\
Other (e.g., American Indian, Pacific Islander)  & 0.9\% & 2\%\\
\end{tabular}
\caption{Gender, age, education, income, race/ethnicity compositions among participants of the main survey ($n$$=$$1,386$). Census statistics from \protect\cite{census2022quick,census2022education,census2022income,census2022age} as of 2022. Some percentages do not add up to 100\% due to non-reporting.}~\label{tab:pmt-participants}
\end{table*}


Table~\ref{tab:pmt-participants} summarizes the demographics of our main survey participants compared to the US census bureau's data. Our sample has a quite balanced distribution of gender, income, and race/ethnicity, but it is slightly more educated and younger than the US population. 
A minority of participants reported having studied or worked in computer science/information technology (314; 22.7\%), and fewer reported having studied or practiced law or other legal services (49; 3.5\%).

We asked participants how they used the email account they provided for our study. Most participants used the email for an extended period (mean: 12.81 years, median: 12). Most participants checked the email daily (85.4\%); the rest checked the email weekly (11.7\%), monthly or less frequently (2.9\%). In addition, participants could choose multiple options for what they used the email for. Most participants selected using the email account for personal correspondence (86.4\%), followed by signing up for medium-value accounts (75.0\%), signing up for sensitive accounts (57.4\%), signing up for low-value accounts (53.1\%), and professional correspondence (40.8\%). These results indicate that participants were checking breach records for email accounts that they used regularly and for important purposes, which adds to our findings' ecological validity.

\paragraph{Overview of breaches.} We showed each participant information about a real-world password breach associated with their email address, randomly selected from their breach records provided by HIBP.
Our sample consists of 127 unique breaches. The most frequently shown breaches came from Zynga (141), MyFitnessPal (128), MySpace (65), and Chegg (62). Figure~\ref{fig:breach-companies-bubble} in Appendix~\ref{app:breaches} gives a more complete overview of breaches in our sample. Aside from email addresses and passwords, which were exposed in every breach due to our study design, the other most commonly breached data types included names (63), IP addresses (54), and dates of birth (35). For other types of data that got leaked in our sample's breaches, see Figure~\ref{fig:breach-data-types} in Appendix~\ref{app:breaches}.
We also calculated breach age, defined as the time between the breach's occurrence date and the survey's completion date. The average age of breaches in our sample was 5.16 years (median: 4.2, sd: 3.13).

\subsection{Comparisons Between Experimental Conditions}

A majority of participants (868, 62.6\%) in the main survey stated they intended to change the breached password. However, only about a third (320, 27.2\%) reported having changed the password in the follow-up survey. The substantial drop from intention to action suggests that there is an intention-behavior gap in individuals' reactions toward the advice about changing breached passwords, and our qualitative results in Section~\ref{subsec:qual} provide insights into why.

Looking at the descriptive statistics, we observed the following trends: \textit{threat only} $>$ \textit{coping only} $\geq$ \textit{threat + coping} $>$ \textit{control} for motivating password intention; \textit{threat + coping} $>$ \textit{threat only} $\geq$ \textit{coping only} $>$ \textit{control} for motivating actual password changes. However, the differences between groups were minor, which might explain why for the two omnibus $\chi^2$ tests of independence we conducted on intention and action respectively, both tests led to non-significant results ($\chi^2(3)$$=$$6.10$, $p$$=$$.11$ for intention; $\chi^2(3)$$=$$5.27$, $p$$=$$.15$ for action). As such, we rejected both H1 and H2: there was no significant association between the conditions participants were assigned to and whether or not they intended to change, or actually changed their breached passwords.

While the omnibus $\chi^2$ test results were insignificant, we ran pairwise $\chi^2$ tests of independence to test the confirmatory directional hypotheses we developed about the differences between the control and treatment conditions. Significant differences were found for specific pairs, summarized in Tables~\ref{tab:pmt-intention} and~\ref{tab:pmt-action}.
Participants who saw the threat appeal alone were 1.48x more likely to intend to change their breached password than the control condition ($p$$=$$.02$), confirming H1a. Participants who saw both the threat and coping appeals were 1.54x more likely to change their breached password than the control condition ($p$$=$$.02$), confirming H2c. The odds ratio for both comparisons only corresponds to a small effect size~\cite{chen2010big}. We did not find a statistically significant difference between \textit{coping only} vs. \textit{control} for intention ($p$$=$$.23$), between \textit{threat + coping} vs. \textit{control} for intention ($p$$=$$.30$), between \textit{threat only} vs. \textit{control} for action ($p$$=$$.14$), or between \textit{coping only} vs. \textit{control} for action ($p$$=$$.23$). H1b, H1c, H2a, H2b are rejected. These results show that, compared to the control condition, threat appeal alone was more effective at increasing password change intention. However, participants were 
the most likely to act 
only when both threat and coping appeals were present.  

As part of our exploratory analyses, we ran pairwise $\chi^2$ tests between the \rev{three treatment conditions}.
For intention, we did not observe any significant differences between \textit{threat only} vs. \textit{threat + coping} ($p$$=$$.20$), \textit{coping only} vs. \textit{threat + coping} ($p$$=$$.92$), or \textit{threat only} vs. \textit{coping only} ($p$$=$$.27$). The same pattern of non-significant pairwise difference also applies to action: $p$$=$$.46$ for \textit{threat only} vs. \textit{threat + coping}, $p$$=$$.31$ for \textit{coping only}vs. \textit{threat + coping}, and $p$$=$$.85$ for \textit{threat only} vs. \textit{coping only}. These results confirm the pattern in Tables~\ref{tab:pmt-intention} and~\ref{tab:pmt-action}: the differences in motivating password changes between the three treatment conditions were minor (1\% to 4\% based on descriptive statistics) and were not significant even with a large enough sample to detect a small-to-medium effect. 

{\renewcommand{\arraystretch}{1.5}
\begin{table}[tp]\centering
\footnotesize
\begin{tabular}{p{0.25\linewidth}p{0.15\linewidth}p{0.05\linewidth}p{0.10\linewidth}p{0.08\linewidth}} \toprule
\textbf{Conditions} &\textbf{\% w/ intention} & \textbf{OR} & \textbf{95\% CI} & \textbf{p-value} \\ \hline
\rowcolor{lightgray}
Threat only vs. Control &  67.3\% vs. 58.2\% & 1.48 & [1.07, 2.04] & .02 \\ 
Coping only vs. Control &  62.9\% vs. 58.2\% & 1.22 & [0.89, 1.68] & .23 \\ 
Threat $+$ Coping vs. Control & 62.3\% vs. 58.2\% & 1.19 & [0.88, 1.63] & .30 \\ 
\bottomrule
\end{tabular}
\caption{Pairwise comparisons between control vs. treatment for the percent of participants who reported intending to change the password in the main survey.}
\label{tab:pmt-intention}
\end{table}}

{\renewcommand{\arraystretch}{1.5}
\begin{table}[tp]\centering
\footnotesize
\begin{tabular}{p{0.25\linewidth}p{0.15\linewidth}p{0.05\linewidth}p{0.10\linewidth}p{0.08\linewidth}} \toprule
\textbf{Conditions} &\textbf{\% w/ action} & \textbf{OR} & \textbf{95\% CI} & \textbf{p-value} \\ 
\hline
Threat only vs. Control & 28.0\% vs. 22.7\% & 1.32 & [0.90, 1.95] & .14 \\ 
Coping only vs. Control & 27.3\% vs. 22.7\% & 1.29 & [0.86, 1.90] & .23 \\ 
\rowcolor{lightgray}
Threat $+$ Coping vs. Control & 31.1\% vs. 22.7\% & 1.54 & [1.04, 2.27]& .02 \\ 
\bottomrule
\end{tabular}
\caption{Pairwise comparisons between control vs. treatment for the percent of participants who reported having changed the password in the follow-up survey.}
\label{tab:pmt-action}
\end{table}}

\subsection{Manipulation Checks}


Table~\ref{tab:pmt-construct} in Appendix~\ref{app:manipulation-check} shows the descriptive statistics of participants' ratings of the five PMT constructs: threat severity, threat vulnerability, response efficacy, self-efficacy, and response costs. To understand to what extent our nudges produced the intended effect on participants' threat and coping perceptions, we conducted Kruskal-Wallis tests to detect whether participants' ratings of each construct differed between conditions, followed by post hoc Dunn tests to detect significant differences between any pairs. We applied Holm-Bonferroni correction to the post hoc Dunn tests to control for Type I error due to the exploratory nature of the analysis.

Results of the Kruskal-Wallis tests showed that our nudges had different effects on participants' perceived threat vulnerability ($H(3)$$=$$14.11$, $p$$=$$.002$) and perceived response efficacy ($H(3)$$=$$21.52$, $p$$<$$.001$). The post hoc Dunn tests further revealed that participants who received the threat appeal had significantly higher ratings of perceived threat vulnerability, evidenced by the significant differences between threat-only vs. control ($p$$=$$.01$) and threat$+$coping vs. control ($p$$=$$.004$). Interestingly, the threat appeal also seemed to increase participants' coping perceptions. Participants who received the threat appeal had significantly higher ratings of response efficacy, evidenced by the significant differences between \textit{threat only} vs. \textit{control} ($p$$=$$.009$), \textit{threat + coping} vs. \textit{control} ($p$$<$$.001$), and \textit{threat + coping} vs. \textit{coping only}  ($p$$=$$.009$). There was no significant difference between conditions for participants' ratings of threat severity ($H(3)$$=$$0.97$, $p$$=$$.81$), self-efficacy ($H(3)$$=$$1.30$, $p$$=$$.73$), or response costs ($H(3)$$=$$4.06$, $p$$=$$.26$); the post hoc Dunn tests did not reveal any significant pairwise differences for these three variables.

These results suggest that our nudges' manipulations of PMT constructs were somewhat but not fully successful. The threat appeal primarily influenced participants' threat perception by influencing perceived threat vulnerability. In addition, the threat appeal's presence strengthened participants' perception of response efficacy, even though response efficacy is part of coping appraisal according to PMT, whereas the presence of coping appeal alone did not help much. This could mean that our coping appeal did not produce desired effects on participants' coping appraisals. Another possible explanation is that particular challenges associated with changing breached passwords make a coping appeal inapplicable --- some participants could not find an account with the breached site to start with; other participants tried the ``forget my password'' feature but never received a password reset confirmation email. We provide more details about these challenges in Section~\ref{subsec:qual}.


\subsection{Other Factors Related to the Breach and Affected Account}
\label{subsec:covariate}

In the main survey, we asked participants to report their awareness of the company/breach, security attitudes, prior negative incidents, as well as how they used the online account affected by the breach. We treat these factors as covariates that could impact their password change behaviors in addition to the condition they were assigned to.

\paragraph{Aware of the company, but not the breach.} For prior awareness (Q\ref{q:prior-site}, Q\ref{q:prior-breach}), the majority (75.8\%) of participants had heard of the breached site before our study, 17.1\% had not, and 7.1\% were unsure. By contrast, the majority (82.3\%) of participants were unaware that they were affected by the breach we showed them before our study, 7.0\% had known they were affected, and 10.8\% were unsure.

\paragraph{Most affected accounts were rarely used and perceived as unimportant.} We further asked participants whether they had an account with the breached site (Q\ref{q:account-existence}), hypothesizing that password change would only be actionable, and thus more likely, for those with an account.\footnote{In Q\ref{q:account-info} and Q\ref{q:account-info-other} we also asked participants to select what types of information the account might have about themselves. However, only in the middle of data collection did we find out that Q\ref{q:account-info} did not have a ``none of the above'' option, and participants could not skip this question. As such, we refrain from reporting the results of this question since the responses were likely skewed due to the survey question's design.} More than half of participants indicated they had created an account with the breached site, either by providing an email address and a password (48.5\%) or by using a social login feature such as ``sign in with Google'' or ``sign in with Facebook'' (6.9\%). The rest did not think they had an account (16.6\%) or were unsure about the account's existence (28.1\%). 

For participants who reported having an account with the breached site, we followed up with questions about the account's age (Q\ref{q:account-age}), frequency of use (Q\ref{q:account-freq}), and perceived importance (Q\ref{q:account-importance}). Most accounts had existed for an extended period (mean: 5.98 years; median: 5; sd: 4.36). Most participants (87.2\%) logged into the account only yearly or even less frequently; the rest checked the account monthly (8.0\%), weekly (3.5\%), or daily (1.3\%). When asked about the account's perceived importance, most participants indicated that the account was ``very unimportant'' (51.1\%) or ``unimportant'' (23.1\%) to them. Average ratings of the perceived importance level were low (mean: 2.04; median: 1; sd: 1.44 for a 7-point scale), \rev{which might be due to a non-trivial amount of participants not knowing whether they had an account with the breached site}. Altogether, these results illustrate the nature of accounts affected by HIBP breaches among our sample: most participants created the account a while ago, only used it very infrequently, and did not attach much value to the account.


\paragraph{Low rate of password reuse for affected accounts.} Excluding 11.0\% of participants who claimed they did not have an account with the breached site, 12.3\% knew for sure they used the breached password elsewhere, 32.5\% did not reuse the breached password, and 44.2\% were unsure. \rev{The low rate of participants who reused the breached password for other accounts matches the finding of most affected accounts being unimportant, as prior work has found that users tend to reuse passwords that they have to enter frequently~\cite{wash2016understanding}.} 

\paragraph{Other covariates.} We examined participants' security attitudes using SA-6~\cite{faklaris2019self} (Q\ref{q:SA6}) and prior experience with account compromise (Q\ref{q:unauth}), data breach (Q\ref{q:noti-breach}), and identity theft (Q\ref{q:id-theft}) \rev{as background variables that do not relate to any specific breach or online account, but could still impact participants' overall security posture}. Most participants' SA-6 ratings were between the medium to high end (mean: 3.44; median: 3.75; sd: 0.95), indicating that most participants were fairly motivated to learn about security and follow expert-recommended advice. More than half of the participants knew that their information was exposed in other data breaches before our study (57.2\% yes; 36.1\% no; 6.6\% unsure). Fewer participants had experienced account compromises (29.1\% yes; 54.6\% no; 16.3\% unsure) or identity theft (12.0\% yes; 81.4\% no; 6.6\% unsure).

\subsection{Regression Results}
\label{subsec:regression}

Having understood participants' awareness of the breach and usage of the affected account through descriptive statistics, we next ran logistic regressions to identify how the effect of the nudges compares to these covariates we measured. \rev{We included the same set of independent variables for the intention model (Table~\ref{tab:pmt-intention-reg}) and action model (Table~\ref{tab:pmt-action-reg}): experimental condition, prior knowledge of the breached site, prior knowledge of the breach, account existence, reuse of the breached password, security attitudes (SA-6), prior negative experience (with account compromise, data breach, and identity theft), demographics (age, gender, education, income), and the breach's timing.}

\begin{table}[htp]
\centering
\footnotesize
\resizebox{0.5\linewidth}{!}{
\begin{tabular}{l c c c c }
 & {\bf B (SE)} & {\bf OR} & {\bf 95\% CI} & {\bf p-value}  \\
\toprule
\rowcolor{lightgray}
(Intercept)   & $-2.88 (0.59)$  & $0.06$ & $[0.02, 0.18]$ &  $<.001$ \\
\midrule
\shortstack{Condition: coping \\ \scriptsize (vs. control)}  & $0.28 (0.24)$ & $1.32$ & $[0.83, 2.11]$ &  $.24$  \\
\midrule
\rowcolor{lightgray}
\shortstack{Condition: threat \\ \scriptsize (vs. control)}  & $0.61 (0.24)$ & $1.84$ & $[1.15, 2.95]$ &  $.01$  \\
\midrule
\shortstack{Condition: combined \\ \scriptsize (vs. control)}  & $0.28 (0.29)$ & $1.32$ & $[0.82, 2.17]$ &  $.25$  \\
\midrule
\shortstack{Account exist: yes \\ \scriptsize (vs. no)}  & $-0.15 (0.32)$ & $0.86$ & $[0.45, 1.60]$ &  $.63$  \\
\midrule
\shortstack{Account exist: yes \\ \scriptsize (vs. no)}  & $-0.28 (0.30)$ & $0.75$ & $[0.42, 1.37]$ &  $.35$  \\
\midrule
\shortstack{Aware account: yes \\ \scriptsize (vs. no)}  & $0.50 (0.35)$ & $1.64$ & $[0.82, 3.29]$ &  $.16$  \\
\midrule
\shortstack{Aware account: unsure \\ \scriptsize (vs. no)}  & $0.15 (0.34)$ & $1.17$ & $[0.59, 2.28]$ &  $.65$  \\
\midrule
\rowcolor{lightgray}
\shortstack{Password reuse: yes \\ \scriptsize (vs. no)}  & $1.10 (0.31)$ & $3.01$ & $[1.66, 5.69]$ &  $<.001$  \\
\midrule
\rowcolor{lightgray}
\shortstack{Password reuse: unsure \\ \scriptsize (vs. no)}  & $0.60 (0.19)$ & $1.82$ & $[1.25, 2.67]$ &  $.002$  \\
\midrule
\rowcolor{lightgray}
\shortstack{Security attitudes \\ \scriptsize (5-point scale)}  & $0.65 (0.10)$ & $1.92$ & $[1.58, 2.34]$ &  $<.001$  \\
\midrule
\shortstack{Acc. Compromise: yes \\ \scriptsize (vs. no)}  & $0.15 (0.20)$ & $1.16$ & $[0.79, 1.71]$ &  $.44$  \\
\midrule
\shortstack{Prior breach: yes \\ \scriptsize (vs. no)}  & $-0.24 (0.19)$ & $0.78$ & $[0.54, 1.13]$ &  $.20$  \\
\midrule
\shortstack{Identity theft: yes \\ \scriptsize (vs. no)}  & $0.12 (0.29)$ & $1.13$ & $[0.64, 2.04]$ &  $.68$  \\
\midrule
\shortstack{Age: 35-54 \\ \scriptsize (vs. 18-34)}  & $0.29 (0.20)$ & $1.33$ & $[0.90, 1.96]$ &  $.24$  \\
\midrule
\shortstack{Age: 55+ \\ \scriptsize (vs. 18-34)}  & $0.62 (0.26)$ & $1.87$ & $[1.12, 3.15]$ &  $.15$  \\
\midrule
\rowcolor{lightgray}
\shortstack{Gender: women \\ \scriptsize (vs. men)}  & $0.68 (0.18)$ & $1.97$ & $[1.38, 2.82]$ &  $<.001$  \\
\midrule
\shortstack{Edu.: $\geq$Bach. \\ \scriptsize (vs. $\leq$high school)}  & $0.28 (0.30)$ & $1.32$ & $[0.72, 2.30]$ &  $.36$  \\
\midrule
\shortstack{Edu.: Some College \\ \scriptsize (vs. $\leq$high school)}  & $0.47 (0.30)$ & $1.60$ & $[0.88, 2.90]$ &  $.12$  \\
\midrule
\rowcolor{lightgray}
\shortstack{Income: 100$+$K \\ \scriptsize (vs. $<$50K)}  & $0.52 (0.25)$ & $1.69$ & $[1.05, 2.76]$ &  $.03$  \\
\midrule
\shortstack{Income: 50-100K \\ \scriptsize (vs. $<$50K)}  & $0.03 (0.20)$ & $1.03$ & $[0.69, 1.52]$ &  $.89$  \\
\midrule
\rowcolor{lightgray}
\shortstack{Breach age \\ \scriptsize (years)}  & $-0.09 (0.03)$ & $0.92$ & $[0.87, 0.97]$ &  $.001$  \\
\bottomrule
\end{tabular}}
\caption{Logistic regression for predicting password change intention in the main survey. $n$$=$$725$ after excluding incomplete or not applicable responses. Cox and Snell $R^2$$=$$0.14$. Model $\chi^2(21)$$=$$111.05$, $p$$<$$.001$}\label{tab:pmt-intention-reg}
\end{table}



\begin{table}[htp!]
\centering
\footnotesize
\resizebox{0.5\linewidth}{!}{
\begin{tabular}{l c c c c }
 & {\bf B (SE)} & {\bf OR} & {\bf 95\% CI} & {\bf p-value}  \\
\toprule
\rowcolor{lightgray}
(Intercept)   & $-2.98 (0.68)$  & $0.05$ & $[0.01, 0.19$ &  $<.001$ \\
\midrule
\shortstack{Condition: coping \\ \scriptsize (vs. control)}  & $0.46 (0.28)$ & $1.59$ & $[0.92, 2.74]$ &  $.10$  \\
\midrule
\rowcolor{lightgray}
\shortstack{Condition: threat \\ \scriptsize (vs. control)}  & $0.60 (0.26)$ & $1.83$ & $[1.10, 3.09]$ &  $.02$  \\
\midrule
\rowcolor{lightgray}
\shortstack{Condition: combined \\ \scriptsize (vs. control)}  & $0.78 (0.28)$ & $2.18$ & $[1.26, 3.83]$ &  $.006$  \\
\midrule
\shortstack{Aware site: yes \\ \scriptsize (vs. no)}  & $0.18 (0.36)$ & $1.20$ & $[0.59, 2.47]$ &  $.62$  \\
\midrule
\shortstack{Aware breach: yes \\ \scriptsize (vs. no)}  & $0.22 (0.32)$ & $1.24$ & $[0.66, 2.31]$ &  $.49$  \\
\midrule
\shortstack{Account exist: yes \\ \scriptsize (vs. no)}  & $0.42 (0.40)$ & $1.52$ & $[0.70, 3.42]$ &  $.30$  \\
\midrule
\shortstack{Account exist: unsure \\ \scriptsize (vs. no)}  & $0.17 (0.40)$ & $1.05$ & $[0.69, 1.58]$ &  $.66$  \\
\midrule
\shortstack{Password reuse: yes \\ \scriptsize (vs. no)}  & $-0.03 (0.30)$ & $0.97$ & $[0.53, 1.75]$ &  $.93$  \\
\midrule
\shortstack{Password reuse: unsure \\ \scriptsize (vs. no)}  & $0.05 (0.21)$ & $1.05$ & $[0.69, 1.58]$ &  $.83$  \\
\midrule
\rowcolor{lightgray}
\shortstack{Security attitudes \\ \scriptsize (5-point scale)}  & $0.46 (0.11)$ & $1.59$ & $[1.28, 1.99]$ &  $<.001$  \\
\midrule
\shortstack{Acc. Compromise: yes \\ \scriptsize (vs. no)}  & $-0.10 (0.21)$ & $0.90$ & $[0.59, 1.37]$ &  $.63$  \\
\midrule
\shortstack{Prior breach: yes \\ \scriptsize (vs. no)}  & $-0.05 (0.20)$ & $0.95$ & $[0.64, 1.43]$ &  $.82$  \\
\midrule
\shortstack{Identity theft: yes \\ \scriptsize (vs. no)}  & $0.36 (0.29)$ & $1.44$ & $[0.81, 2.52]$ &  $.21$  \\
\midrule
\shortstack{Age: 35-54 \\ \scriptsize (vs. 18-34)}  & $-0.37 (0.22)$ & $0.69$ & $[0.44, 1.06]$ &  $.09$  \\
\midrule
\shortstack{Age: 55+ \\ \scriptsize (vs. 18-34)}  & $0.33 (0.27)$ & $1.39$ & $[0.83, 2.34]$ &  $.21$  \\
\midrule
\shortstack{Gender: women \\ \scriptsize (vs. men)}  & $-0.06 (0.19)$ & $0.94$ & $[0.64, 1.37]$ &  $.75$  \\
\midrule
\shortstack{Edu.: $\geq$Bach. \\ \scriptsize (vs. $\leq$high school)}  & $0.21 (0.33)$ & $1.23$ & $[0.65, 2.38]$ &  $.53$  \\
\midrule
\shortstack{Edu.: Some College \\ \scriptsize (vs. $\leq$high school)}  & $0.20 (0.33)$ & $1.22$ & $[0.65, 2.34]$ &  $.55$  \\
\midrule
\shortstack{Income: 100$+$K \\ \scriptsize (vs. $<$50K)}  & $0.10 (0.25)$ & $1.11$ & $[0.68, 1.80]$ &  $.68$  \\
\midrule
\shortstack{Income: 50-100K \\ \scriptsize (vs. $<$50K)}  & $-0.01 (0.23)$ & $0.99$ & $[0.63, 1.54]$ &  $.96$  \\
\midrule
\rowcolor{lightgray}
\shortstack{Breach age \\ \scriptsize (years)}  & $-0.11 (0.03)$ & $0.89$ & $[0.83, 0.95]$ &  $<.001$  \\
\bottomrule
\end{tabular}}
\caption{Logistic regression for predicting password change action in the follow-up survey. $n$$=$$615$ after excluding incomplete or not applicable responses. Cox and Snell $R^2$$=$$0.21$. Model $\chi^2(22)$$=$$141.59$, $p$$<$$.001$}\label{tab:pmt-action-reg}
\end{table}




\paragraph{Treatment vs. control differences hold after controlling for covariates.} \rev{We found that participants who saw the threat appeal were still more likely to form password change intention than those in the control condition ($OR$$=$$1.84$, $p$$=$$.01$). Similarly, the significant difference between the \textit{threat + coping} condition and \textit{control} condition still holds for password change action ($OR$$=$$2.18$, $p$$=$$.006$). The regression results also suggest that participants in the \textit{threat only} condition were more likely to end up changing their passwords compared to the control condition ($OR$$=$$1.83$, $p$$=$$.02$). However, because the $\chi^2$ pairwise comparison does not yield significant results, the difference between the two is inconclusive.}


\paragraph{Proactive security attitudes contribute to changing passwords.} \rev{Participants who held more proactive security attitudes, reflected by higher SA-6 ratings, were significantly more likely to both form password change intentions ($\beta$$=$$0.65$, $p$$<$$.001$) and end up changing their passwords ($\beta$$=$$0.46$, $p$$<$$.001$). Our findings contradict some other studies in which SA-6 did not explain variances in their outcome variables, such as using secure mobile payments~\cite{story2020intent} and adhering to privacy setting suggestions~\cite{krsek2022self}. A possible explanation is that some of the SA-6 scale items (e.g., ``I am extremely motivated to take all the steps needed to keep my online data and accounts safe'') are more directly relevant to our study's context.}

\paragraph{Higher likelihood to change passwords for recent breaches.} \rev{Breach age, i.e., how long ago the breach happened in years, emerged as a significant predictor in both models. If the breach happened a while ago, participants were less likely to form password change intention ($\beta$$=$$-0.09$, $p$$=$$.001$) or to end up changing their password ($\beta$$=$$-0.11$, $p$$<$$.001$). This finding also aligns with prior research on consumer reactions to data breaches in which the occurrence of actual harms was often used as a heuristic for deciding whether to take action~\cite{zou2018consumers,mayer2023awareness,abramova2023anatomy}: if a breach happened a while ago and nothing happened, consumers may take it as a signal that no action is needed, although the objective risk of leaked data being misused does not decrease over time.}

\paragraph{Predictors for password change intention but not action.} \rev{Some predictors, namely whether the breached password was reused elsewhere and the participant's gender and income, could significantly explain the variances in participants' password change intention. Participants who reused the breached password for other online accounts, as well as those who were unsure about the password being reused or not, were significantly more likely to form password change intentions ($OR$$=$$3.01$, $p$$<$$.001$; $OR$$=$$1.82$, $p$$=$$.002$). Participants who identified as women were more likely to form password change intentions than men ($OR$$=$$1.97$, $p$$<$$.001$). Participants with a $>$100K annual household income were more likely to form password change intentions than those earning $<$50K ($OR$$=$$1.69$, $p$$=$$.03$). However, these differences no longer held for the action model, and we were unable to conclude confidently about password reuse and demographics being significant predictors.}

\subsection{How Participants Changed Their Passwords}

We asked participants in the follow-up survey ($n$$=$$1,175$) whether they changed the password for other online accounts (Q\ref{q:pmt-other-acc}). For those who indicated they changed the breached password ($n$$=$$319$), we further asked how soon they changed the password since taking the main survey (Q\ref{q:pmt-change-days}), what they used for the new password (Q\ref{q:pmt-new-pass}), and what techniques they used for remembering the new password (Q\ref{q:pmt-how-remember}) to characterize their password change behaviors, using survey questions from prior work~\cite{golla2018site,mayer2022users}. 

\paragraph{Password changes selectively made to other accounts.} For password changes regarding other accounts, almost half of the participants (48.9\%) left passwords for other accounts the same. Fewer participants prioritized changing the password for other accounts using the same or similar passwords (18.5\%) or for accounts they thought were important, such as bank accounts (16.3\%). Only 6.1\% of participants said they changed the password for every online account. The remaining 10.2\% described their action for other accounts via free text, mostly mentioning that they had the habit of changing passwords regularly (e.g., \textit{``I often change all my online passwords at least once or twice yearly''}) or they changed the password for other accounts affected by breaches after checking HIBP (e.g., \textit{``I already use different passwords for most accounts, but I did check the list of hacked sites and I think I changed the password on one''}). Our findings are similar to results by Golla et al.~\cite{golla2018site} in that participants developed their priorities in changing passwords for other accounts and rarely changed passwords across all accounts; however, in our study, even more participants exclusively focused on changing the password for the account affected by the breach. 

\paragraph{Password changes happened promptly.}
Most of our participants who changed their password did it very promptly after receiving our nudges in the main survey (mean: 1.46 days; median: 1; sd: 2.24). Compared to prior work, our participants reacted to the password change prompt much faster---e.g., the mean time taken to reset the password was 26.3 days in Huh et al.'s case study of the LinkedIn breach~\cite{huh2017linkedin}, and only 13\% of participants in Bhagavatula et al.'s study changed their passwords within three months of the breach announcement~\cite{bhagavatula2020how}. The differences could come from the study's methodology: we followed up with our participants two weeks since they received the breach notification, whereas in Huh et al.'s work the time gap between the breach date and data collection was several months~\cite{huh2017linkedin}; we proactively reached out to our participants to ask about their password change, whereas Bhagavatula et al. derived their findings based on naturalistic observational data~\cite{bhagavatula2020how}. The finding implies that our nudges effectively prompted participants to follow through with their intention, and participants who forgot or did not have intention were likely to leave their password as it was without additional reminders.

\paragraph{New passwords were mostly secure.}
About half of our participants (50.2\%) reported using a password completely unrelated to the old one they created. Some participants (32.9\%) used a unique, random password generated by a password manager. Fewer participants exposed themselves to risks of password reuse attacks in creating the new password, either by only changing a few characters in the old password (8.8\%) or by using a password that they already used for other accounts (4.7\%). The remaining 3.4\% self-described their new password, such as following their own password creation heuristics (e.g., \textit{``the same password schema I use for other sites, which is known to me but generates a different password for different sites''}) and using a random password created by themselves without trying to remember it (e.g., \textit{``I won't use that account again so I just typed in a lengthy string of numbers. If I want it again I will just recover the password''}). Compared to findings in Golla et al.~\cite{golla2018site}, our participants had stronger new passwords and much fewer reused their old passwords. This is likely because our participants were creating passwords for their own accounts in real life rather than in a hypothetical scenario. The message in our coping appeal (Figure~\ref{fig:coping-text}) could have also helped as it specifically discouraged password reuse and linked a guideline for making strong passwords.

Participants could choose multiple options to indicate their strategies for remembering the new password. Participants' strategies were diverse; the most popular options were saving the new password in the browser (27.3\%), remembering the new password without writing it down or storing it digitally (25.4\%), using a third-party password manager such as 1Password and LastPass (21.0\%), and writing the new password down on paper (17.9\%). Less commonly, participants stored the new password in a digital file (12.2\%), used a system-provided password manager such as iOS Keychain (10.3\%), or planned to reset the password every time they logged in rather than remember it (2.5\%). Our findings align with Mayer et al.~\cite{mayer2022users} and Pearman et al.~\cite{pearman2019why}: participants' strategies for password management were primarily using a password manager and trying to remember it mentally.

\paragraph{Difficulty in validating self-reported behaviors.} 
We asked participants who indicated they had changed their passwords to optionally upload a screenshot of the password reset confirmation email to verify the validity of their responses. We adapted the method from Huh et al.~\cite{huh2017linkedin} by making the question optional but having a \$1.00 bonus payment to incentivize participants and honor their time. We also explicitly reminded participants to double-check that the screenshot did not include sensitive or personal information in our instructions (Appendix~\ref{q:pmt-screenshot}). Only 77 (24.1\%) participants uploaded a screenshot; the rest either could not find the email (127, 39.8\%) or chose not to upload a screenshot (115, 36.1\%). In open-ended responses, participants mainly explained that they had deleted the email permanently (\textit{``I deleted the email after receiving it, and it has probably been cleared from my Trash folder too since I try not to make my inbox too cluttered''}), they did not think it was worth the effort (\textit{``I don't feel like searching through my email right now. Even though I would like the extra payment''}), or they were uncomfortable with the question (\textit{``I don't feel comfortable sending my information to someone I don't personally know''}). Due to the low response rate, we did not use this question as a validation mechanism for participants' self-reported password change behaviors. While the question being optional could contribute to the low-response rate, we believed that it was a more ethical approach so that participants were not put in a position to sacrifice their data for monetary gains unwillingly. The different findings between our study and Huh et al.'s study~\cite{huh2017linkedin} could also imply shifting norms among crowdworkers as they become more attentive and protective of their privacy~\cite{sannon2022privacy}.

\section{Qualitative Results}
\label{subsec:qual}
\mr{
We analyzed participants' open-ended responses regarding their rationales for changing or not changing breached passwords (Q\ref{q:pmt-intention} and Q\ref{q:pmt-action}), as well as their general reactions after learning about the breach (Q\ref{q:pmt-response}). Table~\ref{tab:pmt-combined-reason} summarizes the most prevalent themes, which we unpack below.


{\footnotesize
\begin{table}[tp]
\centering
\begin{adjustbox}{width=0.8\linewidth,keepaspectratio}
\begin{tabular}{l r l r}
\textbf{Password Change: Yes} & \textbf{Count} & \textbf{Password Change: No} & \textbf{Count}\\
\cmidrule(lr){1-2} \cmidrule(lr){3-4}
to be safe  & 319  &   inactive use & 471 \\
bad things  & 284  & no account & 321 \\ 
take other actions  & 222  &  no sensitive info & 166 \\
triggered by breach & 140  &  take other actions  & 120\\
inactive use  &  123  & unimportant account & 109 \\
\end{tabular}
\end{adjustbox}
\caption{Top reasons for changing or not changing the breached password, combing responses to Q\ref{q:pmt-intention} ($n$$=$$1,386$) and Q\ref{q:pmt-action} ($n$$=$$1,175$).}
\label{tab:pmt-combined-reason}
\end{table}}



\paragraph{Changing the password: proactive attitude.}

A common theme among participants who changed their password was a proactive attitude toward staying safe and secure ($n$$=$$319$),\mr{
\footnote{In our findings, we report frequency counts of the most prevalent themes to help the reader understand the relative magnitudes. We consider this a valid approach as our study has a relatively large sample size, and the responses were easy to code and reliably coded. Future research that cites our qualitative findings should be careful about quoting the absolute numbers, as they were tied to a very specific setting and might not generalize well.}} usually in the expression of \textit{``just to be safe,''} \textit{``to have a peace of mind,''} and \textit{``better safe than sorry.''} Such a proactive attitude could drive the action even when the participant thought the risk was limited or the account was unimportant; as one explained, \textit{``I'm not too concerned about my MySpace info being hacked as it was such a long time ago \dots I will however change my password if that is an option just to be on the safe side.''} 

\paragraph{Changing the password: fear of negative consequences.} Participants also mentioned various ``bad things'' that could occur as a result of being affected by the breach ($n$$=$$284$), which could reflect the effect of our threat appeal. 
One participant detailed how their personal information could be misused, \textit{``My LiveJournal account contains fiction I wrote that I never want to lose. The knowledge that someone could delete all that work is quite perturbing.''} Another participant mentioned concerns related to credential stuffing attacks, 
\textit{`` I want to change it because I use a lot of recycled passwords. Therefore, criminals could access other, more important accounts like e-mail accounts, bank accounts, my Amazon account, and other online retailer accounts.''} This quote also aligns our previous quantitative finding that participants who reused, or were unsure about reusing the leaked password elsewhere were more likely to form password change intention.

\paragraph{Not changing the password: account inactive or low-value.}
For participants who did not (intend to) change the password, the account no longer being used was the top reason ($n$$=$$471$). Relatedly, a substantial portion of participants also commented that the account did not contain much sensitive information ($n$$=$$166$) or was unimportant ($n$$=$$109$), indicating that they did not perceive the account to be high-value. As one participant reasoned, \textit{``This account isn't important, and I'm pretty sure I haven't logged into it in almost a decade. The information contained in it would be minimal since I never really shared personal details or provided accurate information to websites for certain questions.''} The unimportant comment sometimes also applies to the breached password, as one participant said, \textit{``the password I used for it is an old and way too simple password that I don't use for my important accounts. I have used that password for some accounts like free online games but that is it.''}

\paragraph{Not changing the password: no account with the breached site.} Participants also commented on the practical constraints that made password change impossible or difficult. For instance, 321 participants believed that they did not have an account with the breached site, as they did not recognize the site's name, recognized the site but did not remember making an account, or the account was already deleted. Without an account, the advice of changing the breached password for that account becomes unrealistic, as one participant wrote, \textit{``I do not have a Quidd account so there is no password to change.''}
Another 86 participants confirmed that their account indeed no longer existed, as they tried to log into the account but failed, because the site did not recognize the email address provided, the site was down, or they never received the password reset email. 
Data brokers might also contribute to the problem, as one participant speculated, \textit{``When I tried to reset my password, the email never came \dots Maybe these sites had bought my information somehow from some other place, and therefore were able to obtain a username/email and password from me despite me never opening an account with them directly.''}

{\footnotesize
\begin{table}[tp]
\centering
\begin{adjustbox}{width=0.45\linewidth,keepaspectratio}
\begin{tabular}{l r}
\textbf{Other Actions} & \textbf{Count}\\
\cmidrule(lr){1-2}
change other passwords  & 171 (27.1\%) \\
delete this account  & 89 (14.1\%) \\
try accessing account but fail  & 89 (14.1\%) \\
check breach records  & 59 (9.3\%) \\
check reused passwords  & 48 (7.6\%) \\
check/delete info in account  & 35 (5.5\%) \\
check out the breached site & 32 (5.1\%) \\
review other accounts  & 27 (4.3\%) \\
\end{tabular}
\end{adjustbox}
\caption{Additional actions participants took after learning about the breach, based on responses to Q\ref{q:pmt-response} ($n$$=$$632$).}
\label{tab:pmt-other-action}
\end{table}}

\paragraph{Actions taken other than changing the password.}
As ``take other actions'' emerged as a common theme, regardless of whether the participant changed the breached password, we summarize what these actions are in Table~\ref{tab:pmt-other-action}, derived from responses to Q\ref{q:pmt-response}. 
When changing the password for other online accounts ($n$$=$$171$), some participants made the change to all accounts 
while others prioritized important accounts, e.g., \textit{``I changed my Gmail password since that is the only account I'm particularly worried about.''}  To determine which accounts were at higher risk, 59 participants obtained a more comprehensive list of breach records to guide their password changes.
Another 48 participants reviewed their existing passwords to identify those that were reused and should be changed first. 


Some other actions taken by participants were unrelated to passwords. For instance, 89 participants deleting their account on the breached site, as one participant recounted, \textit{``After the previous survey, I did change my password. Later that day, I decided to just delete my account. I never really did use that site anyway, and it just got me thinking that I really didn't need to have any info at all on it.''} 
Expecting that they may need to use the account in the future, another 35 participants checked content in the account to ensure there was no sensitive information. As one participant detailed their experience, \textit{``I immediately went to my LiveJournal account, where I changed my password and the email address associated with the account. I also confirmed that all posts which contained potentially identifying information were either deleted or set to private.''} Other actions taken include searching for more information about the breached site ($n$$=$$32$) and reviewing and sometimes deleting other inactive accounts ($n$$=$$27$).


}
\section{Discussion}

Our experiment compared the effectiveness of nudges that incorporate a threat appeal, a coping appeal, or both, against a control condition. We focused on the impacts of these nudges on participants' intended and actual behaviors around password changes after data breaches. \rev{Compared to the control condition, a threat appeal alone significantly increased participants' intention to change their passwords, and threat and coping appeals combined significantly increased participants' password change behavior. The effect size was small in both comparisons, and no significant differences were found between the three treatment conditions.}
Participants further described issues they encountered when trying to change their breached password and alternative actions they took. We next discuss our study's limitations, then summarize how our findings contribute new knowledge to PMT literature and their implications for designing compromised credential notifications.


\subsection{Limitations}
Our work has multiple limitations. First, we situated our nudges in breach records curated by HIBP. While HIBP is a reputable source and has been integrated into other services such as Firefox Monitor and 1Password, the database mainly consists of username and password breaches since it is built on scans of account credential dumps. Other types of personally identifiable information, such as social security numbers and medical records, rarely appear in HIBP's database, but they do appear in other databases of breach records~\cite{itrc2022annual}. Some of our findings  
could be different if participants were presented with breaches that exposed more intimate details of themselves. \rev{Because publicly available APIs for accessing breach records (like the one provided by HIBP) are still rare,} future research could build a tool to curate breach records from multiple sources or partner with companies and/or non-profit organizations to replicate our results. 


Second, \rev{some of the limitations relate to our sampling of participants}. We conducted our experiment only with participants recruited in the US. 
We relied on Prolific for recruitment, which led to the exclusion of less tech-savvy participants such as those without Internet access. These recruitment criteria limit our findings' generalizability beyond the US and more tech-savvy consumers. Thus, they introduce opportunities for future research to replicate our study in different locations with different legal and consumer protection frameworks around data breaches.

\rev{Third, our evaluation of nudges  relies on participants' self-reported data. While self-reported data is subject to issues such as recall and social desirability biases, prior research has shown self-reported data varies consistently and systematically with measured data in security user studies when the experimental manipulations have non-trivial variations~\cite{redmiles2018asking}, which is the case in our study. Our attempt to validate participants' self-reported action via screenshots was unsuccessful due to the low response rate for this optional question. Future research could implement our experimental protocol on real-world measurement data and compare them to self-reported data to identify whether a significant gap exists between the two.}

Lastly, a minority of our survey questions were not perfect measures for their intended constructs. As discussed in Section~\ref{subsec:covariate}, we had to exclude responses about the account's information type due to the absence of a ``none of the above'' option. Additionally, our questions about the account's age, the purpose of use, and perceived importance might not capture all nuances as participants' account usage evolves.\footnote{One participant wrote, ``I had a hard time answering one question: `To the best of your memory, how long have you been using your LiveJournal account?' I used it all through college and then forgot about it for 13+ years, until just now. So I used it for a few years, but it's not true that I've `used it for a few years,' but it's definitely also not true that I've used it for 13+ years?'' Unfortunately, we got this response in the middle of data collection, so we could not make further changes to the question.} 
As lessons learned, future work could consider specifying the timing of account-checking (e.g., ask ``When was the last time you checked the account?'' rather than ``How often do you check the account?'') and using Likert scales rather than single Likert items for concepts such as perceived account importance to more accurately characterize users' diverse account usage behaviors.

\subsection{Theoretical Contributions}


\paragraph{Comparing our findings to existing PMT literature.}

Our findings add to the ongoing discussion in PMT research about whether threat or coping appraisal plays a more prominent role in individuals' formation of protection motivation. Our findings highlight an interesting pattern: compared to the control condition, the threat appeal alone performed significantly better at raising password change intention, but the combination of threat and coping appeals performed significantly better at driving actual password changes. 

\mr{Comparing our findings with PMT's application in other security contexts, our findings confirm that it usually requires both threat and coping appeals to drive action, as has been shown in the cases of adopting secure mobile payments~\cite{story2020intent} and Tor browser adoption~\cite{story2022increasing}. Comparing our findings to those of van Bavel et al.~\cite{van2019using}, who similarly differentiated threat from coping appeals, we observe a different pattern, as in their study the coping appeal was more effective than the threat appeal in nudging participants toward secure online purchases~\cite{van2019using}. The different finding might be due to the simulated environment in their experiment: participants would not have experienced the errors and usability issues that can occur with real-world online accounts, which could impact their coping behaviors subsequently.}

\mr{Comparing our findings with PMT's application in other contexts beyond security and privacy also highlights the uniqueness of our study's context and how it might shape the findings. In two meta-analyses of the PMT literature primarily in the health domain, the coping appraisal components were found to have greater predictive validity than the threat appraisal components for both behavioral intentions and actual behaviors~\cite{milne2006prediction,floyd2000meta}. The different patterns could be due to the different context featured. In health-education interventions, the threat that the protective behavior seeks to address---whether it is stroke, heart disease, or cancer---is likely to be obvious and salient to most people. By comparison, the threat of a breached password may be more nebulous. Nevertheless, highlighting this threat, especially as we walked participants through how credential stuffing attacks work, could also make a bigger difference. Our context also introduces unique considerations for the coping behavior. While a health-related behavior (e.g., breast self-examination, smoking cessation, or adopting a healthy diet) is applicable to most people, this is not the case for changing a breached password. As evidenced by our qualitative results, a non-trivial portion of our participants did not have an account with the breached site or encountered issues as they tried to change the breached password. These are unfortunate scenarios that cannot be addressed by a coping appeal, no matter how effective it is.}

\paragraph{The limitations of applying PMT to motivate password changes.}

Our results shed light on the limitations of PMT by identifying other factors that explain variances in participants' password change behavior and the hurdles participants experience in changing their passwords. \rev{An example of such factors is attitude.} Prior work has showcased how attitude influenced security intention and behavior~\cite{ifinedo2012understanding,siponen2014employees,van2019using}, yet `attitude' is a broad term that can apply to many subjects, from the recommended action to risks in general. Our work highlights the importance of security attitudes more precisely: \rev{from the regression results, SA-6 was a significant predictor in both the intention and action models; from the qualitative findings, ``to be safe'' is a prevalent theme.}

Another contextual factor that impacts password change behaviors is the breach's timing. The regression results suggest that participants were less likely to change the account password for breaches that happened a long time ago. Our qualitative analysis further provides insights into why. For example, one participant reasoned that so much time has passed that the breached information was no longer relevant, \textit{``The breach happened so long ago that I'm sure that password is not being used anywhere else for my accounts online. Along with the other compromised information, it's no longer the same, except my name which is common knowledge and published on the internet anyways.''} 


Moreover, while PMT exclusively focuses on appraisals and hurdles in human cognitive processing, what needs to be considered in our study's particular context is systemic issues regarding the password ecosystem.
For instance, forgetting or not knowing an old password was a common theme for inaction, yet when an average American user has over 150 online accounts that require a password~\cite{dashlane2018account}, keeping track of different passwords for every single account is fundamentally challenging, especially for those who have not developed a password management system~\cite{pearman2017let,florencio2007large}. In attempting to change their passwords, participants encountered some issues that were technically out of their control, e.g., the site told them there was no account that matched the provided email address, they struggled to find the password reset button, or they struggled to identify which account to change when the breached company owned many sites that the account could belong to. Some of these issues might be unintended consequences of the site's attempt to comply with privacy regulations, e.g., when the site deletes inactive accounts to meet GDPR's data minimization requirement~\cite{gdpr2017}. Others were akin to the usability issues of privacy controls~\cite{habib2019empirical,habib2020usability}, and could even be viewed as dark patterns that sites deploy to coerce users into staying in business with them~\cite{mulligan2020fertile}.

\subsection{Practical Implications}

\rev{Our nudges have several promising avenues for deployment. Breach notifications are already mandated by many data breach disclosure laws around the world~\cite{romanosky2011data,zou2019you}, mostly via mailed letters but also increasingly via digital channels such as emails.}
For password breaches in particular, compromised credential checking had been adopted as a feature in mainstream browsers (e.g., Google Chrome and Microsoft Edge)~\cite{pullman2019protect,lauter2021password} as well as in password managers (e.g., 1Password and LastPass)~\cite{shiner2022finding,fremery2021breaking}. \rev{Another way to notify users is during the next time they access a particular online account that has been affected by breaches, as this represents a natural moment for them to interact with the breached password.}
Our findings help inform best practices for designing notifications in these settings, as we discuss below.


\paragraph{Highlight threats, but with caution.}

Prior work has highlighted misunderstandings users may have in assessing the risks of data breaches and password reuse~\cite{zou2018consumers,mayer2021now,bhagavatula2020how,golla2018site} as well as proposed ideas to address them. For example, Golla et al.~\cite{golla2018site} advocated that the notification should encourage users to change similar passwords on other accounts and thoroughly explain why doing so mitigates password reuse attacks. Huang et al.~\cite{huang2022users} drew specific recommendations for Chrome's compromised credential notification, suggesting more explanations about why it is necessary to change the breached password and what risks the user may face if they do not change their passwords. Our study directly tested Huang et al.'s proposed idea, and our findings confirm that a threat appeal is useful to raise users' intention to change their breached credentials.

Nevertheless, the threat appeal alone performed only marginally better than the control condition at raising password change intentions, if judged by the effect size. As such, we caution that service providers should consider the returns and potential unintended consequences of overly emphasizing threats. 
One needs to consider whether iterating, implementing, and evaluating the threat appeal is worth the effort if a plain notification with less text can achieve the same purpose most of the time. To avoid overwhelming users, prior work has recommended providing information in a layered form~\cite{huang2022users,emami2019exploring,cranor2012necessary}, and we incorporated this format in our visual displays of the nudging text. Unfortunately, we cannot know if the layered approach truly results in an improvement since we did not conduct A/B testing on this feature. This could be an opportunity for future research, especially considering that the specific layer designs will likely differ across products and interfaces. 

Furthermore, there might be inevitable tensions between highlighting threats and making the notification trauma-informed. Security incidents like data breaches can trigger traumatic and stressful experiences~\cite{chen2022trauma}. Prior work has highlighted that a threat appeal, when being too strong, could be counter-productive, especially for those already in a vulnerable state: they may engage in risk denial or simply refuse to engage with the fearful message as a self-protective mechanism because the threat makes them feel uncomfortable~\cite{ruiter2014sixty}. Our participants appreciated our nudges for informing them of the breach, and no one explicitly commented on the message being too triggering. Yet some participants recounted how busy they were or how they got distracted by other things,\footnote{One participant mentioned other duties they were struggling to deal with in life, \textit{I have saved the information, but I was ill and I had to work three days 12 hours shift in a row, so I haven't had a chance to do anything.''} Another participant shared that the breached site reminded them of family members who had passed away, \textit{``If this is true then it would have been my son who used it on my phone when his was broken. My son is now dead (suicide), so I cannot ask him.''}} suggesting that an inappropriately designed threat appeal could generate unnecessary burdens or anxiety. This is where a coping appeal could come into play and ease potential stress reactions triggered by the threat appeal, such as by telling the user that they could pick a later time to deal with the issue and asking them if they want a reminder for this~\cite{story2022increasing}.

\paragraph{Communicate data flows to mitigate distrust.} 

Trust issues can prevent users from adopting password managers or adhering to security advice. In Huang et al.'s study~\cite{huang2022users}, many participants' concerns were related to Google, as they assumed that Google checks their non-saved credentials, or they worried about Google taking control over their own data. By contrast, none of our participants expressed concerns about HIBP, which we explicitly highlighted as the source of the breach records, perhaps because of the service's non-profit nature. Interestingly, our participants' concerns mostly centered around the breached site, e.g., when they were unfamiliar with the site's name 
or when they were concerned about additional data leaks that could happen as they sought to recover their account.

These findings highlight the necessity of providing more transparency to mitigate potential questions, distrust, and misunderstandings among users, but how to do so well remains a key challenge. Prior work has recommended that browsers and password managers that send compromised credential notifications should explain how the product detects leaked or reused passwords~\cite{huang2022users} and, on a higher level, what measures the company takes to protect users' data~\cite{redmiles2019should}. Service providers are already doing this to some extent in their messaging (e.g., see ``Privacy is at the heart of our design'' in Google's blog post about its password checkup feature)~\cite{pullman2019protect}. Still, the deeper issue is users' lack of awareness of and trust in encryption~\cite{wu2018tree,dechand2019encryption}, which will likely cause comprehension problems for such messaging. Helping users build trust with the breached site is more tricky, as users have reasonable doubts about the site for leaking or exposing their data in the first place. Proactively reaching out to affected users might help, as our participants expressed that they would rather learn about and deal with the breach rather than stay uninformed. For instance, one participant said, \textit{``I do not remember receiving any notice from Poshmark. They can't even spend the resources to notify us.''} If the user trusts the site enough to check it out, there could be additional explanations of when and how their account was created (if it still exists) or why their account no longer exists.

\paragraph{Standardizing and semi-automating the password change experience.}

The various challenges our participants experienced in attempting to change their passwords highlight larger problems with the password ecosystem. As much as we tried to predict edge cases in developing our coping appeal, 
our findings reflect that the real-world situation is far more complex.     Participants could encounter all kinds of scenarios in which things could go wrong. It then becomes difficult to predict all scenarios and enumerate corresponding solutions in a concise coping appeal.

Our findings reflect that the experience of changing a password could be drastically different from site to site. However, it should not be this way, and we see the need for standardizing the password change processes by providing more industry guidelines or pushing for stronger regulations. For example, the California Consumer Privacy Act (CCPA) requires businesses to include a ``Do Not Sell My Personal Information'' link on their website that allows consumers to submit an opt-out request (if the business sells personal information)~\cite{ccpa2020final}. We imagine similar efforts could be made to standardize where websites should provide the account login fields as well as information regarding how to change the account password. The CAN-SPAM Act in the US requires that for commercial messages, businesses should honor consumers' opt-out requests within ten business days~\cite{ftc2009canspam}. Similarly, we see opportunities for standardizing the turnaround time for sending password reset confirmations, given that one of the hurdles our participants reported was struggling to find the password reset email.

Beyond standardization, we see opportunities for partially automating the password change experience. For instance, for someone who already adopts a password manager, the password manager could provide a feature that allows the user to scan through every saved login credential and determines whether it still works for the corresponding site.

\paragraph{Opportunities and challenges for providing personalized advice.}

Our findings reveal the cost-benefit analysis behind participants' decision to change the breached password. The action was reasonable for someone who used the account extensively and felt motivated to protect it from potential compromises. However, participants who did not care about the account tended to delete the account instead, and participants who were already using strong and unique passwords for different accounts had valid reasons for not changing the password since their risk levels were minimal. 

Our findings call for a deeper reflection on whether changing the breached password should be the nudging goal for everyone given the diverse risk levels and preferences among individuals.  
We see the promises of personalized advice to help consumers better assess whether password changes are needed in their particular situations. For instance, existing password managers such as 1Password and LastPass are already providing dashboards and scores that help users evaluate overall password strength and the extent of password reuse~\cite{devito2022how,lastpass2022digital}.
Going beyond these features, password managers could provide personalized and data-driven recommendations on which account(s) to prioritize (such as banking and shopping sites that are more likely to contain financial information) in cleaning up old or weak passwords. In addition, advice on which actions to take should ideally be tailored to the specific types of compromised information. 
Breaches that involve more sensitive or unique data types, such as social security numbers or medical records, might require more serious measures. Existing advice for what to do after data breaches mostly enumerates different actions depending on the different breached data types~\cite{johansen2021what}. A more efficient way could be presenting personalized advice in the order of priority. The interactive resource provided by the US Federal Trade Commission~\cite{ftc2020when} is a good step toward this direction, although the extent of personalization there is still quite limited. 

While we see exciting opportunities offered by personalizing post-breach coping advice, we should not ignore challenges associated with personalization. Too much automation, while reducing users' burden, could reduce their sense of agency~\cite{colnago2020informing}. The personalized advice might be inaccurate or irrelevant, which may further cause trust issues or annoy users. When the advice is about password changes in particular, the personalization will likely need to be offered via password managers, which have an increasing yet still small user base~\cite{mayer2022users}. The dependence on password managers might also exclude specific populations like older adults, who are known to have more trust issues with cloud storage of passwords~\cite{ray2021older}. The questions then become: how do we find a balance between making personalized advice useful versus learning too much about users' preferences to the extent of causing privacy concerns? How do we make personalized advice accessible to everyone and truly reflect users' diverse needs and preferences?

\section*{Acknowledgments}

The authors thank Denise Anthony, Tawanna Dillahunt, Byron Lowens, Elissa Redmiles, and Peter Story for their valuable feedback on study design and early drafts. This research was partially supported by a University of Michigan Rackham Graduate Student Research Grant.



\bibliographystyle{plain}
\bibliography{ref}

\appendix
\section{Survey Material}
\label{app:pmt-survey-material}

\newcounter{qcounter}
\newcommand{\mychoice}[1]{{$\circ$} #1 \, }

\subsection{Screening Survey}
\label{app:screening-survey}

\subsubsection*{Informed consent}

\textit{[This informed consent form was for the screening survey. We used a very similar consent form for the main and follow-up surveys, except adjusting the estimated completion time and compensation.]}

\noindent {\bf Study Title:} Consumers' Reactions Toward Data Breaches

\noindent {\bf Principal Investigators:} {\em REDACTED}

\noindent{\bf Purpose of this Study:}
We are conducting a research study to understand how consumers perceive and react to data breaches.

\noindent{\bf Description of your involvement:} 

If you agree to be part of the research study, we will ask you to complete a screening survey that asks you to provide an email address. We will query this email address in haveibeenpwned.com, a public database, to see if it has appeared in any data breaches. 

Depending on the query results, we may send you an invitation to our main survey following your completion. The main survey will show more information about these breaches and ask you to answer a few questions about them.

\noindent{\bf Compensation:}

We expect this screening survey to take \textbf{about two to three minutes}. You will be compensated \textbf{\$0.80} upon completing the survey.

You are free to withdraw at any time. However, you will not be compensated if you withdraw from the study.

\noindent{\bf Benefits and Risks:}
Although you may not directly benefit from participating in this study, the study will inform how to better protect consumers against data breaches. 

The risks associated with your participation are similar to those normally encountered when using the Internet. We take strong measures to protect your personal information, as we outline in the ``Confidentiality'' section.

\noindent{\bf Confidentiality:}
By participating in the study, you understand and agree that the {\em REDACTED} may be required to disclose your consent form, data and other personally identifiable information as required by law, regulation, subpoena or court order. 

Otherwise, your confidentiality will be maintained in the following manner:

\begin{itemize}[noitemsep,topsep=0pt]
    \item Your data will be stored in password-protected cloud services and will only be accessible to the study team.
    \item We plan to publish the results of this study, but we will not include any information that would identify you.
    \item Throughout the study, you will not be asked to provide any direct personal identifiers in the study apart from your email address. \textbf{We do not track your email address, and we will not be able to tie your email address to any results or analysis. All records of your email address will reside in temporary storage to facilitate the lookup of data breaches, and will be deleted following the completion of this task. We will never see your actual email address.}
\end{itemize}

\noindent{\bf Right to Ask Questions \& Contact Information:}
If you have questions about this research, you may contact the study team at {\em REDACTED}.

The {\em REDACTED} Institutional Review Board has determined that this study is exempt from IRB oversight.

\noindent{\bf Voluntary Consent:}
By proceeding to the next page, you are agreeing to participate in this study. You may print a copy of this consent form for your records. You may also contact the study team at any time by emailing {\em REDACTED} if you think of a question later.

Are you 18 years of age or older? \mychoice{Yes} \mychoice{No}

Are you physically living in the United States? \mychoice{Yes} \mychoice{No}

\textit{[Participants need to answer ``Yes'' for both questions to proceed to the next page. If the answer is ``No'' to any question, show an error message ``We're sorry, but to participate in this survey you must be at least 18 and currently physically located in the United States. Thank you for your interest.'']}

\subsubsection*{Breach lookup}

We are going to ask you to enter your most commonly used email address at the bottom of this page. We will use your email address to look up whether your email address has appeared in any data breaches (also called ``security breaches''), using the public lookup service for data breaches haveibeenpwned.com. Based on the results, we may invite you to our main survey in which we will show you more information about these breaches.

{\bf Privacy Notice:} \textbf{We do not track or store your email address as part of this study, and we will not be able tie your email address to any results or analysis.} All records of your email address will reside in temporary storage to facilitate the lookup of data breaches, and will be deleted following the completion of this task. We will never see your actual email address.

To access information about breaches, your email address will be communicated to haveibeenpwned.com, a public service that maintains a database of data breaches involving email addresses. Communication with haveibeenpwned.com will occur on secure and encrypted channels. haveibeenpwned.com does not permanently store email addresses used in queries as described in their privacy policy.

If you have any further concerns about providing your email address, you may opt-out of the survey at this time. We will remove any record of your participation. Note that if you choose to opt out, you will not be compensated.

\begin{enumerate}[noitemsep,topsep=0pt]
    \item Please enter your most commonly used email address. After the task, you may search for another email address, but for now, we are primarily interested in breaches that may have involved your most commonly used email address. [free text]
\setcounter{qcounter}{\value{enumi}}
\end{enumerate}

\subsubsection*{Email-related questions}

Please tell us more about this email address.

\begin{enumerate}[noitemsep,topsep=0pt]
\setcounter{enumi}{\value{qcounter}}
\item Whose email address is it?
    \mychoice{It is my own account / I have sole ownership of this account}
    \mychoice{It is my shared account / I share the account with someone else (e.g., a partner or family member)}
    \mychoice{It is someone else's account / someone else has sole ownership of this account}
    \mychoice{I made up an email address just for this study}

\item How often do you check emails in this account?     
    \mychoice{Every day}
    \mychoice{A few times a week}
    \mychoice{A few times a month}
    \mychoice{A few times a year or less frequently}

\item What do you use this email account for? Choose all that apply. 
    \mychoice{For personal correspondence (e.g., friends and family members)}
    \mychoice{For professional correspondence (e.g., with colleagues, business partners)}
    \mychoice{Sign up for sensitive accounts (e.g., banking, taxes)}
    \mychoice{Sign up for medium sensitive accounts (e.g., social media, online shopping)}
    \mychoice{Sign up for low-value accounts (I used it when I'm prompted to sign up but don’t really care)}
    \mychoice{Other [free-text]}

\item How long have you been using this email account? [number entry] 
    \mychoice{year(s)}
    \mychoice{month(s)}
    \mychoice{week(s)}
    \mychoice{day(s)}
    
\item How many other email addresses/accounts do you regularly use? (Not counting the one you entered) [number entry]

\item Prior to our study, have you ever checked if this email address has appeared in any data breaches using Have I Been Pwned (haveibeenpwned.com) or other services? 
    \mychoice{Yes}
    \mychoice{No}
    \mychoice{Unsure}
    
\setcounter{qcounter}{\value{enumi}}
\end{enumerate}

\subsubsection*{Demographics-related questions}

As the final part of this survey, please tell us a few things about yourself.

\begin{enumerate}[noitemsep,topsep=0pt]
\setcounter{enumi}{\value{qcounter}}
\item What is your age? [number entry or ``prefer not to say'']

    \item What is your gender? 
        \mychoice{Man}
        \mychoice{Woman}
        \mychoice{Non-Binary}
        \mychoice{Prefer to self-describe: [free text]}
        \mychoice{Prefer not to say}

    \item What is the highest level of education you have completed? 
        \mychoice{Less than high school}
        \mychoice{High school or equivalent}
        \mychoice{Some college, no degree}
        \mychoice{Associate's degree, occupational}
        \mychoice{Associate's degree, academic}
        \mychoice{Bachelor's Degree}
        \mychoice{Master's Degree}
        \mychoice{Professional degree}
        \mychoice{Doctoral degree}
        \mychoice{Prefer not to say}

    \item Have you studied or worked in the field of computer science or information technology?
        \mychoice{Yes}
        \mychoice{No}
        \mychoice{Prefer not to answer}

    \item Have you studied or practiced law or other legal services?
        \mychoice{Yes}
        \mychoice{No}
        \mychoice{Prefer not to answer}
        
    \item What was your total household income before taxes during the past 12 months?
        \mychoice{Under \$15,000}
        \mychoice{\$15,000 to \$24,999}
        \mychoice{\$25,000 to \$34,999}
        \mychoice{\$35,000 to \$49,999}
        \mychoice{\$50,000 to \$74,999}
        \mychoice{\$75,000 to \$99,999}
        \mychoice{\$100,000 to \$149,999}
        \mychoice{\$150,000 or above}
        \mychoice{Prefer not to say}
        
    \item Which of the following best describes you? Choose one or more.
        \mychoice{White}
        \mychoice{Black or African American}
        \mychoice{Asian}
        \mychoice{American Indian or Alaska Native}
        \mychoice{Native Hawaiian or other Pacific Islander}
        \mychoice{Middle Eastern or North African descent}
        \mychoice{Prefer to self-describe: [free text]}
    
    \item Are you Hispanic or Latino?
        \mychoice{Yes}
        \mychoice{No}
        \mychoice{Prefer not to say}
\setcounter{qcounter}{\value{enumi}}
\end{enumerate}

\subsubsection*{Next steps (only for eligible participants)}

We will send you the link to our main survey within the next week. In the main survey, we will show you more details about the data breach records associated with your provided email address.

Please click the ``continue'' button to proceed to the next page, where you will be automatically redirected to Prolific.

Stay tuned, and we look forward to seeing you back!

\subsubsection*{Debrief (only for ineligible participants)}
\textit{[For participants who did not have any password breach but had other data breaches associated with their provided email address:]}

Thank you for completing our screening survey. We're looking for participants who have at least one password breach (i.e., a data breach that exposes users' account passwords) to take our main survey. According to haveibeenpwned.com, your email address has not appeared in any password breach.

That being said, your email address has appeared in the following data breaches, and we suggest that you take necessary precautions. Please note that you can always obtain the same results by checking your email address on haveibeenpwned.com, which, in addition, provides records with sensitive breaches upon the verification of your email account. Please keep in mind that this list only reflects breaches that are registered in the haveibeenpwned.com database, your information may have been exposed in other breaches.

\textit{[For participants who did not have any data breach with their provided email address:]}

Thank you for completing our screening survey. We're looking for participants who have at least one password breach (i.e., a data breach that exposes users' account passwords) to take our main survey. According to haveibeenpwned.com, your email address has not appeared in any data breach.

That's great news! However, we still recommend that you monitor breach services like haveibeenpwned.com in case new breaches come to light. In case you still want to learn about what to do when you're affected by a breach, you can find several resources below.

\textit{[For all ineligible participants:]}

Below is a list of resources to help you better protect yourself from data breaches.
\begin{itemize}[noitemsep,nolistsep]
\item Resources about recovering from a data breach:
\begin{itemize}[noitemsep,nolistsep]
    \item Federal Trade Commission: Identity theft recovery steps
    \item Federal Trade Commission: Credit Freeze FAQs
    \item Firefox Monitor: What to do after a data breach
    \item Norton: What to do after 5 types of data breaches
\end{itemize}

\item Resources about protecting yourself against future breaches:
\begin{itemize}[noitemsep,nolistsep]
    \item Firefox Monitor: How to create strong passwords
    \item Firefox Monitor: Steps to protect your online identity
\end{itemize}

\end{itemize}

\subsection{Main Survey}
\label{app:main-survey}

\subsubsection*{Show breach}

In our previous survey, you provided an email address for querying the Have I Been Pwned (haveibeenpwned.com, referred to as ``HIBP'' onward) database for data breach records. Below is a data breach in which your provided email address has appeared. \textit{[Show breach information.]}

\subsubsection*{Breach-related questions}

\begin{enumerate}[noitemsep,nolistsep]

\setcounter{enumi}{\value{qcounter}}
    \item Prior to our study, have you ever heard of [site name]? 
    \label{q:prior-site}
        \mychoice{Yes} 
        \mychoice{No} 
        \mychoice{Unsure}
    \item Prior to our study, were you aware that you are affected by the [site name] data breach? \label{q:prior-breach}
        \mychoice{Yes} 
        \mychoice{No} 
        \mychoice{Unsure}
    \item To your knowledge, do you have an account with [site name]?
    \label{q:account-existence}
        \mychoice{Yes, I created an account by providing an email address and a password}
        \mychoice{Yes, I created an account through a third-party service (e.g., ``sign in with Google'' or ``sign in with Facebook'')}
        \mychoice{No, I did not have an account with [site name]}
    \item \textit{[If ``unsure'' for Q16/Q17/Q18]} You selected unsure for [Q16/Q17/Q18]. Please explain why you selected ``unsure.'' [free text]
    \item \textit{[If ``yes'' for Q18]} To the best of your memory, how long have you been using your [site name] account? [number entry] 
    \label{q:account-age}
    \mychoice{year(s)}
    \mychoice{month(s)}
    \mychoice{week(s)}
    \mychoice{day(s)}
    \item \textit{[If ``yes'' for Q18]} How often do you use your [site name] account?
    \label{q:account-freq}
    \mychoice{Every day}
    \mychoice{A few times a week}
    \mychoice{A few times a month}
    \mychoice{A few times a year or less frequently}
    \item \textit{[If ``yes'' for Q18]} To the best of your memory, does your [company name] account have any of the following information about you? Please select all that apply.
    \label{q:account-info}
    \mychoice{Date of birth}
    \mychoice{Financial information (e.g., bank accounts or credit card numbers}
    \mychoice{Gender}
    \mychoice{IP address}
    \mychoice{Name}
    \mychoice{Phone number}
    \mychoice{Residential address}
    \item \textit{[If ``yes'' for Q18]} Is there any other type of information the account may have about you? If you cannot think of any, please write ``I don't know.'' [free text]
    \label{q:account-info-other}
    \item \textit{[If ``yes'' for Q18]} How important or unimportant is your [site name] account to you?
    \label{q:account-importance}
    \mychoice{Very unimportant}
    \mychoice{Unimportant}
    \mychoice{Somewhat unimportant}
    \mychoice{Neither important nor unimportant}
    \mychoice{Somewhat important}
    \mychoice{Important}
    \mychoice{Very important}
\setcounter{qcounter}{\value{enumi}}
\end{enumerate}

\subsubsection*{Measuring password change intention}

\textit{[Show intervention, see Figures~\ref{fig:threat-text} and~\ref{fig:coping-text} for the specific text.]}

\begin{enumerate}[noitemsep,nolistsep]
\setcounter{enumi}{\value{qcounter}}
    \item After learning about this breach, do you intend to change the password of your [site name] account?
        \mychoice{Yes, I intend to change the password of my [site name] account}
        \mychoice{No, I do not intend to change the password of my [site name] account}
        \mychoice{I have already changed the password of my [company name] account after this breach occurred (on [date]) and before taking this survey}
    \label{q:pmt-intention-close-ended}
    \item Please explain why you selected this answer option. [free text]
    \label{q:pmt-intention}
    \item Do you use your [site name] account's password for any other online accounts?
    \label{q:pw-reuse}
        \mychoice{Yes}
        \mychoice{No}
        \mychoice{I don't know}
        \mychoice{I don't have an account with [site name]}
\setcounter{qcounter}{\value{enumi}}
\end{enumerate}

\subsubsection*{Measuring PMT constructs}

\textit{[The order of questions in this section was randomized.]}

\begin{enumerate}[noitemsep,nolistsep]
\setcounter{enumi}{\value{qcounter}}

\item Please rate to what extent the following incidents would be a serious problem to you. [Answer options for each: not at all serious, slightly serious, somewhat serious, serious, extremely serious.]
\label{q:threat-severity-measure}
    \mychoice{Experience financial loss}
    \mychoice{Have my personal information sold to marketers}
    \mychoice{Have my online accounts hacked by someone}
    \mychoice{Have my identity stolen by someone}
    \mychoice{Receive more spam emails}
    \mychoice{Please select ``extremely serious'' (this is an attention check)}
\item As a result of the [site name] data breach, how likely or unlikely do you think you are to experience the following incidents? [Answer options for each: very unlikely, somewhat unlikely, neither likely nor unlikely, somewhat likely, very likely.]
\label{q:threat-vulnerability-measure}
    \mychoice{Experience financial loss}
    \mychoice{Have my personal information sold to marketers}
    \mychoice{Have my online accounts hacked by someone}
    \mychoice{Have my identity stolen by someone}
    \mychoice{Receive more spam emails}
    \mychoice{Please select ``very unlikely'' (this is an attention check)}
\item Please rate your level of disagreement or agreement with the following statement: ``Changing the password of my [site name] account will protect me from negative incidents as a result of the [site name] breach.''
\label{q:response-efficacy-measure}
    \mychoice{Strongly disagree}
    \mychoice{Somewhat disagree}
    \mychoice{Neither agree nor disagree}
    \mychoice{Somewhat agree}
    \mychoice{Strongly agree}
\item How easy or difficult do you think it would be for you to change the password of your [site name] account?
\label{q:self-efficacy-measure}
    \mychoice{Very difficult}
    \mychoice{Somewhat difficult}
    \mychoice{Neither easy nor difficult}
    \mychoice{Somewhat easy}
    \mychoice{Very easy}
\item If I were to change the password of my [site name] account, I would... [Answer options for each: not true at all, slightly true, somewhat true, true, extremely true.]
\label{q:response-cost-measure}
    \mychoice{Spend a lot of time changing the password}
    \mychoice{Changing the password requires me to learn new skills}
    \mychoice{Feel more anxious about the [site name] data breach}
    \mychoice{Forget the new password and be locked out of the account}
\setcounter{qcounter}{\value{enumi}}
\end{enumerate}

\subsubsection*{SA-6 \& prior negative experience}

You're almost done! Just a few questions about yourself.

\begin{enumerate}[noitemsep,nolistsep]
\setcounter{enumi}{\value{qcounter}}

\item Please indicate your agreement with the following statements. [Answer options for each: strongly disagree, somewhat disagree, neither agree nor disagree, somewhat agree, strongly agree.]
\label{q:SA6}
    \mychoice{Generally, I diligently follow a routine about security practices.}
    \mychoice{I always pay attention to experts' advice about the steps I need to take to keep my online data and accounts safe.}
    \mychoice{I am extremely knowledgeable about all the steps needed to keep my online data and accounts safe.}
    \mychoice{I am extremely motivated to take all the steps needed to keep my online data and accounts safe.}
    \mychoice{I often am interested in articles about security threats.}
    \mychoice{I seek out opportunities to learn about security measures that are relevant to me.}
    
\item Has anyone ever gained unauthorized access to one of your online accounts? E.g., someone secretly changed your password without you noticing it.
\label{q:unauth}
    \mychoice{Yes}
    \mychoice{No}
    \mychoice{Unsure}
    
\item Have you ever learned that your information was exposed in a data breach before taking our survey?
\label{q:noti-breach}
    \mychoice{Yes}
    \mychoice{No}
    \mychoice{Unsure}

\item Have you ever been a victim of identity theft? E.g., someone secretly applied for a new credit card under your name.
\label{q:id-theft}
    \mychoice{Yes}
    \mychoice{No}
    \mychoice{Unsure}

\item \textit{[If ``unsure'' for Q34/Q35/Q36]} You selected unsure for [Q34/Q35/Q36]. Please explain why you selected ``unsure.'' [free text]
\setcounter{qcounter}{\value{enumi}}
\end{enumerate}

\subsubsection*{Final remarks}

Thank you for completing this survey! Please note that the information about this data breach we showed to you is real. Your email address and potentially other personal information has appeared in this breach and could be used by criminals to steal your identity or access your online accounts.

You can check a full list of data breaches associated with your email address on haveibeenpwned.com, which, in addition, provides records with sensitive breaches upon the verification of your email account. Please keep in mind that this list only reflects breaches that are registered in the haveibeenpwned.com database, your information may have been exposed in other breaches.

Please click the ``continue'' button to proceed to the next page, where you will be automatically redirected to Prolific.

\subsection{Follow-up Survey}
\label{app:follow-up-survey}

\subsubsection*{Reminder of breach}

In our previous survey, you provided an email address for querying the Have I Been Pwned (haveibeenpwned.com, referred to as ``HIBP'' onward) database for data breach records. 

Below is a data breach in which your provided email address has appeared.

\textit{[Show breach information.]}

\begin{enumerate}[noitemsep,nolistsep]
\setcounter{enumi}{\value{qcounter}}
    \item Since taking our previous survey on [date], what did you do, if anything, after learning that your information was exposed in the [site name] data breach? Please explain why. [free text]
        \label{q:pmt-response}
\setcounter{qcounter}{\value{enumi}}
\end{enumerate}

\subsubsection*{Attention check}

\begin{enumerate}[noitemsep,nolistsep]
\setcounter{enumi}{\value{qcounter}}
    \item This is an attention check. What is the name of the company that suffered a data breach and exposed your information, as we show you in the previous page?
        \mychoice{correct answer}
        \mychoice{AKP emails}
        \mychoice{KnownCircle}
        \mychoice{Staminus}
\setcounter{qcounter}{\value{enumi}}
\end{enumerate}

\subsubsection*{Measuring password change behavior}

\begin{enumerate}[noitemsep,nolistsep]
\setcounter{enumi}{\value{qcounter}}
\item You took our previous survey on [date]. The survey showed that your information was exposed in the [site name] data breach.  Since then, have you changed the password for your [site name] account?
\label{q:pmt-action-binary}
    \mychoice{Yes}
    \mychoice{No}
\item Please explain why you changed or did not change your [site name] account's password after taking our previous survey on [date]. [free text]
\label{q:pmt-action}
\item What did you do about passwords for other accounts?
\label{q:pmt-other-acc}
    \mychoice{I changed the password for every online account I have}
    \mychoice{I changed the password for other accounts that use the same or similar passwords}
    \mychoice{I changed the password for really important accounts (e.g., bank accounts)}
    \mychoice{I kept using the same password for other accounts}
\item \textit{[If ``Yes'' for Q40]} To your best estimate, how soon after taking our previous survey on [date] did you change the password for your [site name] account? [number entry] days
\label{q:pmt-change-days}
\item \textit{[If ``Yes'' for Q40]} What did you use for the new password for your [site name] account? 
\label{q:pmt-new-pass}
    \mychoice{A password that I already use for other accounts}
    \mychoice{Something related to the old password but a few characters different, created by myself}
    \mychoice{Something completely unrelated to the old password, created by myself}
    \mychoice{A unique or random password, such as one generated by a password manager}
    \mychoice{Other: [free text]}
\item \textit{[If ``Yes'' for Q40]} What techniques did you use to remember the new password for your [site name] account? Please select all that apply.
\label{q:pmt-how-remember}
    \mychoice{I remembered my new password without writing it down or storing it digitally}
    \mychoice{I reset my password every time I log in rather than remembering my new password}
    \mychoice{I wrote my new password down on paper or other physical media}
    \mychoice{I stored my new password in a digital file or files}
    \mychoice{I saved my new password in the browser (e.g., passwords saved in Chrome)}
    \mychoice{I used a system-provided password manager (e.g., Apple's Keychain)}
    \mychoice{I used a third-party password manager (e.g., 1Password or LastPass)}
    \mychoice{Other: [free text]}
\setcounter{qcounter}{\value{enumi}}
\end{enumerate}

\subsubsection*{Screenshot upload}
\label{q:pmt-screenshot}

\textit{[Only display this page if ``yes'' for Q40.]}

Optional: Please upload a screenshot of the password reset confirmation email you received from [site name] after changing the password. Make sure the image you are uploading is in PNG format. We would greatly appreciate it if you do this, as it will help us validate your responses.

If you upload a valid screenshot upon our verification, you will receive a \$1.00 bonus payment in addition to the \$1.20 base payment. 

Steps to take:
\begin{itemize}[noitemsep,nolistsep]
    \item Sign in to your email account; make sure this account has the email address you checked for breaches in our earlier survey.
    \item Do a keyword search of ``[site name]'' in your inbox and/or spam folder.
    \item On a Windows PC, open the ``Snipping Tool'' program; on a Mac computer, press Shift + Command + 4 on your keyboard to take a screenshot.
    \item Select the area you want to take a screenshot of. Make sure your screenshot includes the email's subject line, sender, and date (see the example below).
    \item \textbf{IMPORTANT: Do NOT upload a screenshot that includes your personal information, such as your actual email address, username, and new password.}
\end{itemize}

\begin{enumerate}[noitemsep,nolistsep]
\setcounter{enumi}{\value{qcounter}}

\item Which of the following options applies to you?
\label{q:pmt-screenshot-question}
    \mychoice{I have my screenshot and I am ready to upload it on the next page.}
    \mychoice{I cannot find the password reset confirmation email (please explain why): [free text]}
    \mychoice{I choose not to upload the screenshot (please explain why): [free text]}
\setcounter{qcounter}{\value{enumi}}
\end{enumerate}

\subsubsection*{Final remarks}

Thank you for your participation! As we noted in our previous surveys, the information about the data breach we showed to you is real. Your email address and potentially other personal information has appeared in this breach and could be used by criminals to steal your identity or access your online accounts.

Below is the full list of breaches associated with the email address you provided. Please note that you can always obtain the same results by checking your email address on haveibeenpwned.com, which, in addition, provides records with sensitive breaches upon the verification of your email account. Please keep in mind that this list only reflects breaches that are registered in the haveibeenpwned.com database, your information may have been exposed in other breaches.

\textit{[Show all breaches.]}

\begin{itemize}[noitemsep,nolistsep]
\item Resources about recovering from a data breach:
\begin{itemize}[noitemsep,nolistsep]
    \item Federal Trade Commission: Identity theft recovery steps
    \item Federal Trade Commission: Credit Freeze FAQs
    \item Firefox Monitor: What to do after a data breach
    \item Norton: What to do after 5 types of data breaches
\end{itemize}

\item Resources about protecting yourself against future breaches:
\begin{itemize}[noitemsep,nolistsep]
    \item Firefox Monitor: How to create strong passwords
    \item Firefox Monitor: Steps to protect your online identity
\end{itemize}
\end{itemize}

We greatly appreciate the insights you contributed, which would help us develop better technologies and interfaces that notify people of data breaches.

Would you like to be contacted by us to participate in our future research? \mychoice{Yes, I am interested.}
\mychoice{No, I am not interested.}

Please click the ``continue'' button to proceed to the next page, where you will be automatically redirected to Prolific.
\section{Cognitive Walkthrough Protocol}
\label{app:walkthrough-protocol}

\subsubsection*{Introduction}

Today we will be testing a series of three surveys for an experiment. To ensure you have a fresh experience, I will not reveal the experiment's purpose, but I am happy to talk about it toward the end. 

The first survey is a screening survey. You will be asked to provide an email address for querying a database and see if this email address has appeared in any data breaches.

The second survey is the main survey. You will see more information about a breach associated with the email address you provided, and answer some questions about it.

The third survey is a follow-up survey. For real-world participants, they will receive a link to this survey two weeks after the second survey. Since we are doing pilot testing today, I will ask you to review the follow-up survey as well.

The link to all surveys is [URL]. If possible, could you please share your screen with me as you complete the surveys? We'll be using a think-aloud protocol, which means you will read out the survey questions on your screen line by line, tell me your answer and why you select it. Please also feel free to comment on your thoughts and reactions to the survey questions, especially if there's anything that's unclear or doesn't make sense to you.

For most of the time, I will be quietly sitting in the background and observe your interactions. I may chime in if there is a critical question I want to ask on the spot. I also have a list of questions I would like to get your feedback on toward the end after you complete all three surveys.

I will be taking notes as I observe your interactions. However, to make sure I capture everything, would you mind me recording our meeting today as well?

\subsubsection*{Questions to ask}

\begin{itemize}[noitemsep,nolistsep]
    \item Does the text under ``What are the risks'' describe the threats clearly? Does the text under ``How to change your password'' provide useful information? Do you feel motivated to change an exposed password after reading the text?
    \item Right now there's a 2-second delay after ``What to do'' and a 10-second delay after the threat/coping module. The delay seeks to nudge participants to pay close attention to the text while waiting. Is the delay too long, too short, or just the right amount for you?
    \item For the follow-up survey, how's your experience of completing the screenshot upload question? Are the instructions clear? Is this question in line with ethical data collection? Are they written in a way that minimizes accidental information leaks from taking screenshots of emails?
    \item Any unclear or confusing wording for any survey questions?
\end{itemize}

\section{Breaches in Our Sample}
\label{app:breaches}
 \setlength{\abovecaptionskip}{-1pt} 
 \setlength{\belowcaptionskip}{-5pt} 

    \begin{figure}[!htb]
  \includegraphics[width=0.85\linewidth]{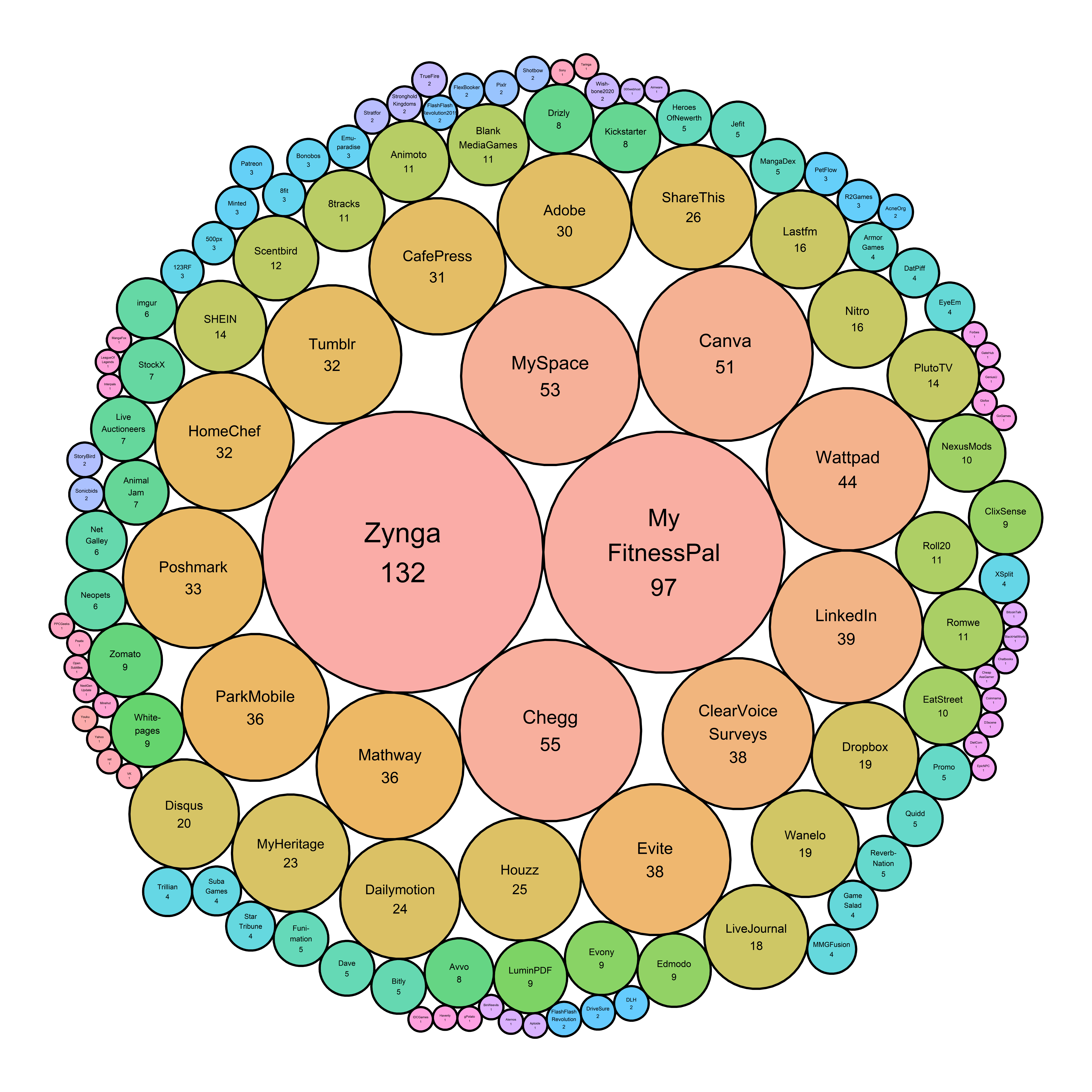}
    \caption{Company name and frequency of the breaches featured in our study (bubble size proportionate to each breach's frequency).}
    \label{fig:breach-companies-bubble}
     
\end{figure}
\begin{figure}[!htb]
  \includegraphics[width=\linewidth]{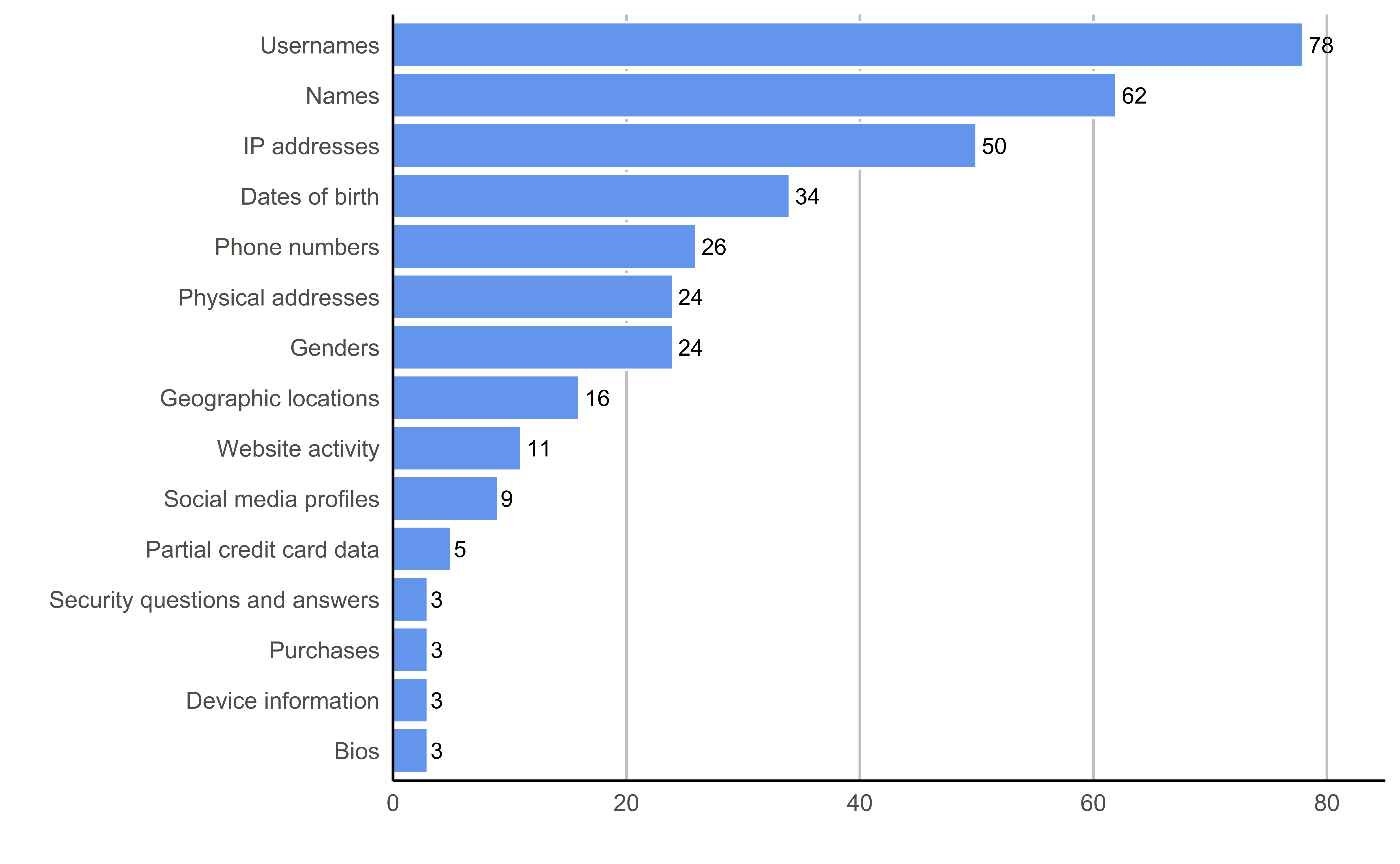}
    \caption{How often each data type got leaked among the 127 breaches in our sample, excluding email addresses and passwords (appearing in all breaches). Eliminated 26 other types occurring twice or fewer.}
    \label{fig:breach-data-types}
\end{figure}

\clearpage 
\section{Manipulation Check Results}
\label{app:manipulation-check}

{\renewcommand{\arraystretch}{1.5}

 \setlength{\abovecaptionskip}{1pt} 
 \setlength{\belowcaptionskip}{5pt} 

\begin{table}[ht]\centering
\footnotesize
\begin{tabular}{p{0.25\linewidth}p{0.20\linewidth}p{0.10\linewidth}p{0.10\linewidth}p{0.10\linewidth}} \toprule
\textbf{Variable} &\textbf{Condition} & \textbf{Mean} & \textbf{Median} & \textbf{SD} \\ \hline
\multirow{4}{*}{Threat severity} &  Control & 4.28 & 5.00 & 0.90  \\ 
 &  Threat only & 4.37 & 5.00 & 0.79  \\ 
 &  Coping only & 4.31 & 4.50 & 0.88  \\ 
 &  Combined & 4.32 & 4.00 & 0.83  \\ 
 \hline
\multirow{4}{*}{Threat vulnerability} &  Control & 2.73 & 3.00 & 0.06  \\ 
 &  Threat only & 2.99 & 3.00 & 0.06  \\ 
 &  Coping only & 2.89 & 3.00 & 0.06  \\ 
 &  Combined & 3.01 & 3.00 & 0.06  \\ 
 \hline
 \multirow{4}{*}{Self-efficacy} &  
 Control & 3.99 & 4.00 & 1.14  \\ 
 &  Threat only & 3.96 & 4.00 & 1.15  \\ 
 &  Coping only & 4.04 & 4.00 & 1.13  \\ 
 &  Combined & 4.00 & 4.00 & 1.07  \\ 
 \hline
\multirow{4}{*}{Response efficacy} &  
 Control & 3.20 & 3.00 & 1.13  \\ 
 &  Threat only & 3.44 & 4.00 & 1.15  \\ 
 &  Coping only & 3.28 & 4.00 & 1.17  \\ 
 &  Combined & 3.55 & 4.00 & 1.13  \\ 
 \hline
\multirow{4}{*}{Response costs} &  
 Control & 1.51 & 1.00 & 0.74  \\ 
 &  Threat only & 1.52 & 1.00 & 0.79  \\ 
 &  Coping only & 1.51 & 1.00 & 0.79  \\ 
 &  Combined & 1.60 & 1.50 & 0.80  \\ 
\bottomrule
\end{tabular}
\caption{Mean, median, and standard deviation of participants' rating of the PMT constructs, divided by conditions. We adapted scales from prior work~\cite{boehmer2015determinants,story2020intent} in measuring these variables: threat severity (Q\ref{q:threat-severity-measure}), threat vulnerability (Q\ref{q:threat-vulnerability-measure}), response costs (Q\ref{q:response-cost-measure}) were measured using 5-point Likert scales, taking the median. Response efficacy (Q\ref{q:response-efficacy-measure}) and self-efficacy (Q\ref{q:self-efficacy-measure}) were measured using a single 5-point Likert-type item.}
\label{tab:pmt-construct}
\end{table}}
\newpage
\section{Qualitative Codebook}
\label{app:codebook}

We provide our codebook for each coded question as well as the respective count for each code.

\subsection{Password Change Intention (Main Survey, Q\ref{q:pmt-intention})}
\begin{itemize*}[noitemsep]
\item \textbf{Yes (868)}:
\textit{to be safe (223), bad things (212), take other actions (181), inactive use (103), triggered by breach (98), action good idea (65), reuse passwords (47), depend on account access (40), just changed Password (40), forget password (24), yes sensitive info (23), other existing measures (19), action easy (18), instructed by survey (18), unspecific (18), action important (17), tried and failed (17), no sensitive info (16), unsure if already changed (16), no account (15), unaware account existence (15), prior negative experience (14), unsure account existence (14), distrust company (13), resigned (11), use unique passwords (9), don't know site (8), important account (8), unsure about info in account (8), forget account existence (7), unimportant password (7), already changed password (6), account future use (6), old breach (5), limited impact (4), account created via social login (3), unimportant email (1)}
\item \textbf{No (518)}
\textit{inactive use (189), no account (152), sensitive info no (80), forget password (45), take other actions (45), unimportant password (43), unimportant account (42), limited impact (29), tried and failed (28), resigned (23), use unique passwords (23), already changed password (21), don't know site (19), too much effort (18), unsure account existence (18), old breach (16), other existing measures (13), don't know how (8), reuse passwords (8), account created via social login (8), unaware account existence (8), unspecific (8), distrust company (7), distrust intervention (5), forget account existence (5), unimportant email (5), prior negative experience (3)}
\end{itemize*}

\subsection{Actions Taken (Follow-up Survey, Q\ref{q:pmt-response})}
\begin{itemize*}[noitemsep]
\item \textbf{Action (632)}:
\textit{change this password (214), change other passwords (171), delete this account (89), try accessing account but fail (89), check breach records (59), check reused passwords (48), check/delete info in account (35), check out the breached site (32), review other accounts (27), check info leaked in this breach (20), enable two-factor authentication (19), other miscellaneous action (11), research on cybersecurity (11), unsubscribed (10), inform others (9), start to use password managers (9), check credit reports (8), use credit monitoring services (8), check financial statements (7), use antivirus (6), use stronger passwords (5), be careful with emails (4), create a new email (4), delete other accounts (4), stay vigilant (4)}
\item \textbf{Nothing (543)}
\end{itemize*}

\subsection{Password Change Behavior (Follow-up Survey, Q\ref{q:pmt-action})}
\begin{itemize*}[noitemsep]
\item \textbf{Yes (855)}:
\textit{to be safe (96), bad things (72), trigger by breach (42), take other actions (41), inactive use (20), to access account (18), instructed by survey (16), reuse passwords (15), action easy (12), action good idea (11), change passwords regularly (10), no sensitive info (10), yes sensitive info (9), distrust company (8), already changed password (7), refer to intervention text (7), use password managers (7), unspecific (6), use unique passwords (5), resigned (3), action important (2), account future use (2), prior negative experience (2), don't know site (1), limited impact (1)}
\item \textbf{No (320)}
\textit{inactive use (282), no account (169), no sensitive info (86), tried and failed (86), forget to do (78), take other actions (75), unimportant account (67), forget password (59), unimportant password (39), busy (38), don't know site (24), limited impact (24), too much effort (24), use unique passwords (20), already changed password (19), action unnecessary (17), unspecific (15), distrust company (14), resigned (12), not a priority (9), old breach (9), account created via social login (8), unsure account existence (3), use password managers (3), change passwords regularly (2)}
\end{itemize*}

\end{document}